\documentclass[%
twocolumn,
prx,
showpacs
]{revtex4-1}

\usepackage[T1]{fontenc}
\usepackage[latin9]{inputenc}
\usepackage{geometry}
\usepackage{float}
\usepackage{amsmath}
\usepackage{amssymb}
\usepackage{graphicx}
\usepackage{epstopdf}
\usepackage[english]{babel}

\begin{document}

\date{\today}
\title{A novel method for calculating relative free energy of similar molecules in two environments}
\author{Asaf Farhi $^{1}$ }
\email{asaffarhi@post.tau.ac.il}
\author{Bipin Singh $^2$}
\affiliation{$^1$ Raymond and Beverly Sackler School of Physics and Astronomy,
Faculty of Exact Sciences,
Tel Aviv University, IL-6997801 Tel Aviv, Israel}

\affiliation{$^2$Center for Computational Natural Sciences and Bioinformatics (CCNSB), International
Institute of Information Technology Hyderabad (IIIT-H), Gachibowli, Hyderabad, 500032,
India}
\begin{abstract}
Calculating relative free energies is a topic of
substantial interest and has many applications including solvation and binding free energies, which are used in computational drug discovery.
However, there remain the challenges of accuracy, simple implementation, robustness and efficiency, which prevent the calculations from being automated and limit their use. 
Here we present an exact and complete decoupling analysis in which the partition functions of the compared systems decompose into the partition functions of the common and different subsystems. This decoupling analysis is applicable to submolecules with coupled degrees of freedom such as the methyl group and to any potential function (including the typical dihedral potentials), enabling to remove less terms in the transformation which results in a more efficient calculation. Then we show mathematically, in the context of partition function decoupling, that the two compared systems can be simulated separately, eliminating the need to design a composite system. We demonstrate the decoupling analysis and the separate transformations in a relative free energy calculation using MD simulations for a general force field and compare to another calculation and to experimental results. We present a unified soft core technique that ensures the monotonicity of the numerically integrated function (analytical proof) which is important for the selection of intermediates. We show mathematically that in this soft core technique the numerically integrated function can be non-steep only when we transform the systems separately, which can simplify the numerical integration. Finally, we show that when the systems have rugged energy landscape they can be equilibrated without introducing another sampling dimension which can also enable to use the simulation results for other free energy calculations.  
\end{abstract}
\medskip
\medskip

\maketitle


\section{Introduction}
Calculating free energy differences between two physical systems,
is a topic of substantial current interest. A variety of advanced
methods and algorithms have been introduced to answer the
challenge, both in the context of molecular dynamics (MD) and Monte
Carlo (MC) simulations 
\cite{allen1987computer,binder2010monte,landauguide,frenkel1996understanding,newman1999monte,chipot2007free,zuckerman2011equilibrium}.
 Applications of these methods include calculations of
binding free energies \cite{kollman1993free,deng2009computations,woods2011water},
free energies of hydration \cite{straatsma1988free}, free energies
of solvation \cite{khavrutskii2010computing}, chemical reactions \cite{jorgensen1987priori} and more.
Free energy methods are extensively
used by various disciplines and the interest in this field is
growing - over 3,500 papers using the most popular free energy
computation approaches were published in the last decade, with the
publication rate increasing $\sim17\%$ per year \cite{chodera2011alchemical}.

Free energy difference between two systems can be calculated using
equilibrium methods (alchemical free energy calculations) and non-equilibrium methods. In equilibrium methods a hybrid system is
used to transform system $A$ into $B$, e.g with the
transformation 
\begin{equation}
\label{eq:first_transformation} 
H_{\mathrm{hybrid}}=\lambda H_{A}+\left(1-\lambda\right)H_{B},\,\lambda \in [0,1],
\end{equation}
(in practice usually more complex transformations are used as will be explained later on). In these methods, the hybrid system is simulated at a set of $\lambda$
intermediates and average values are calculated. Then, using these
values, the free energy difference is calculated. The commonly
used methods include Exponential Averaging/ Free Energy
Perturbation (FEP) \cite{zwanzig1954high} and Thermodynamic Integration (TI)
\cite{frenkel1996understanding,kirkwood1935statistical,straatsma1991multiconfiguration}. Two methods to estimate free energies which are considered equivalent are Bennett Acceptance Ratio (BAR)
\cite{bennett1976efficient} and Weighted Histogram Analysis Method
\cite{kumar1992weighted} (WHAM). 

Another approach which enables to access directly the free energy is the Wang-Landau method, in which random walk is performed in energy space \cite{wang2001efficient,landau2004new}. Other approaches such as $\lambda$ Metadynamics and adiabatic MD \cite{wu2011lambda,abrams2006efficient} suggest to consider $\lambda$ as a coordinate of the system and to enable the system to wander between $\lambda$s. This is performed by introducing in the Hamiltonian a potential term which depends on $\lambda$ which ensures that the system spends more time where the sampling of the free energy as a function of $\lambda$ is more challenging.   

In non-equilibrium methods the work needed in the process of
switching between the two Hamiltonians is measured. These methods
include Jarzynski relation \cite{jarzynski1997nonequilibrium}
and its subsequent generalization by Crooks \cite{crooks2000path}.

Calculating binding free energies is fundamental and has many applications. In particular it has potential to advance
the field of drug discovery which has to cope with new challenges.
In the last years the number of innovative new molecular entities 
for pharmaceutical purposes has remained stable at $5-6$ per
year. This situation is especially grim when taking into account
the continual emergence of drug-resistant strains of viruses and
bacteria. Virtual screening methods, in which the $10^{60}$ possible molecules are filtered out, play a large role in modern drug discovery efforts.
However, there remains the challenge of selecting the candidate molecules out of the
still very large pool of molecules in reasonable times.
Equilibrium methods show great potential in enabling the
computation of binding free energies with reasonable computational
resources. In these methods instead of simulating the
binding processes directly, which would require a simulation many
times the lifetime of the complex, the ligand is transmuted into another through intermediate, possibly non-physical
stages. This is in fact relative free energy calculation in which the difference between free energy of a process of one molecule and another molecule is calculated. 
If the free energy differences between the ligands in the two environments are calculated, the relative binding free energy between the two ligands can be calculated (this cyclic calculation is called the Thermodynamic Cycle).
\begin{figure}[h]
 \centering
\includegraphics[width=8cm]{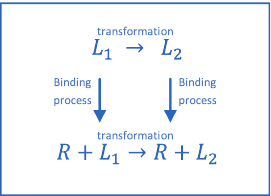}
 \caption{The standard free energy differences scheme in the calculation of binding free energy in the existing methods}
\label{fig:binding}
\end{figure}

In Fig. \ref{fig:binding} a scheme of the free energies in the calculation of binding free energies in the standard methods is presented ($L_1,L_2$ and $R$ represent the ligands and the receptor respectively). For solvation there is a similar scheme in which instead of a receptor there is a solvent.

Free energy calculation methods already have successes in
discovering potent drugs \cite{jorgensen2009efficient}. However, despite the continuing progress in the field from the original concepts,
 the methods have restrictions which prevent them from being automatic and limit their use in
computational drug design.
A naive calculation of the free energy difference using TI can be performed as follows: 
 \begin{align}
\triangle F_{A\rightarrow B}\left(\beta_{1}\right)=\int_0^1\frac{dF_{A\rightarrow B}\left(H_{\mathrm{hybrid}}\left(\lambda\right)\right)}{d\lambda}d\lambda=\\ \nonumber
\int_0^1\int\frac{\left[H_{B}(\mathbf{\Omega})-H_{A}(\mathbf{\Omega})\right]e^{-\beta_{1}\left[\lambda H_{B}(\mathbf{\Omega})+\left(1-\lambda\right)H_{A}(\mathbf{\Omega})\right]}d\mathbf{\Omega}}{Z(\lambda)}d\lambda,
\end{align}
where $\mathbf{\Omega}$ denotes the vector of all coordinates.
It can be seen that at $\lambda=1$ for example $H_{A}$ does not affect the systems' behavior but
its energy values are averaged over, which can result in large magnitudes of the integrated function. Thus, when the systems have low phase space overlap
there are significant changes in the integrated function and large variance and hence large computational cost. This is especially dominant
when the two compared molecules have different covalent bond description which results in a very low phase space overlap (in the naive setup).
Moreover, since molecular force fields include electrostatic and VDW terms that diverge at small atom-atom distances, the average energy can diverge at $\lambda\rightarrow 0,1$.  

\begin{figure}
\includegraphics[width=8cm]{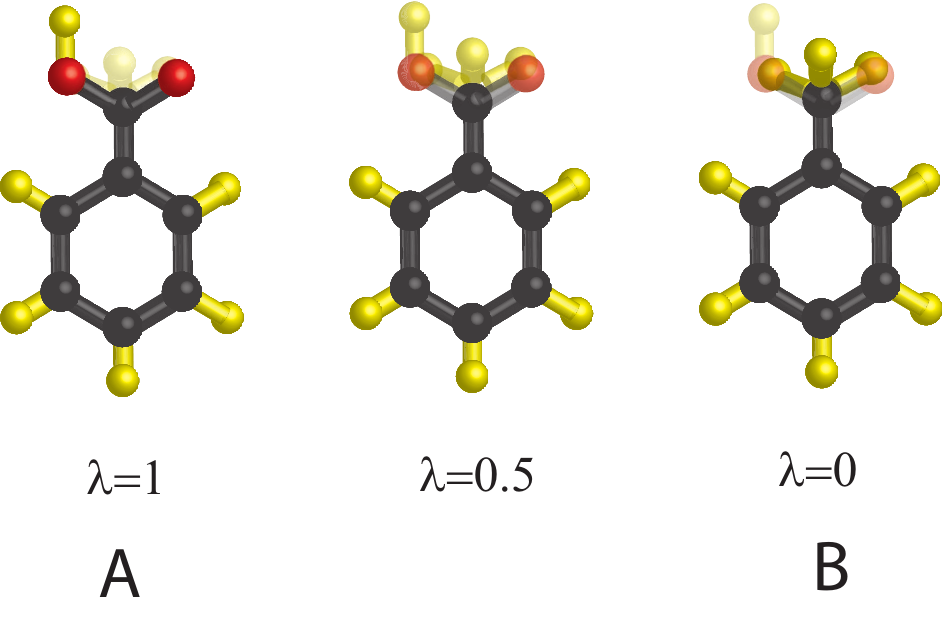}
 \caption{A scheme of the transformation in a hybrid system in the dual topology that compares Benzoic Acid and Toluene at $\lambda$ values of 0,0.5 and 1 (one system). The transparent atoms in the end states $A$ and $B$ are decoupled atoms - atoms whose VDW and Coulomb interactions are removed. At $\lambda=0.5$ the different atoms between the two subsystems are partly interacting. These calculations are often used when the compared systems have a relatively small difference (e.g molecules that differ in few atoms). The transformation represents the top or bottom transformations in Fig. \ref{fig:binding} where $A$ and $B$ represent $L_1$ and $L_2$ respectively.}
 \label{fig:HybridTransformation}
\end{figure}

A variety of approaches and techniques have been introduced to address the challenges in the field. These include the \emph{topologies} for simulating the system, that usually take into account the fact that the compared systems have similarities to generate a hybrid system with higher phase space overlap  (see Fig. \ref{fig:HybridTransformation}). The topologies are usually combined with removing VDW and Coulomb terms of the different atoms which is called \emph{decoupling} in order to further enhance phase space overlap (see e.g Fig. \ref{fig:HybridTransformation}). 
\emph{Soft core} potentials were suggested to avoid singularities at small $\lambda$s. Common sampling techniques to overcome high energy barriers include Temperature and \emph{Hamiltonian Replica Exchange} methods  \cite{ferrenberg1988new,hansmann1997parallel,earl2005parallel}. We will explain these methodologies in the course of the derivation of the method.

However, the calculations in the existing practices have several limitations. First, they are notoriously difficult to implement correctly \cite{shirts2012best}.
Such complications arise for example from the fact that the hybrid system is composed of both systems and hence it usually has to be designed (see for example dual topology in Fig. \ref{fig:HybridTransformation} and Ref. \cite{pearlman1994comparison}).  Moreover, the interactions between atoms from the two compared systems have to be ignored in order for the calculations to be reasonable.
Second,  since the process of transforming one system into the other is different for each
comparison and has no a priori known properties, the choice of intermediates remains a challenge. In the context of TI this is equivalent to a function that needs to be numerically integrated without any known properties.
In addition, since both systems interact simultaneously with the environment the behavior of the intermediate systems cannot be predicted.
Third, each type of hybrid topology has small phase space overlap in one aspect \cite{pearlman1994comparison}. 
Fourth, the existing {decoupling} analysis, while explaining several important principles, involves approximations and does not treat all the potentials \cite{boresch2003absolute} (such as non-quadratic terms, methyl group etc.).
Fifth, the soft core technique, while being efficient in removing singularities from the calculations, has various disadvantages. One of them is difficult implementation due to the complicated functions involved and the requirement to transform first the Coulomb terms and then the VDW terms in order to avoid singularities. In addition, since it involves changing the shape of the functions it results in lower phase space overlap between intermediates. It is worth noting that the free energies associated with the transformations (Fig. 1) are often much larger than their difference. Thus, these calculations have to be very precise to produce reliable relative free energy values.

Temperature Integration (TeI) was suggested in Refs. \cite{FarhiTemperature,FarhiThesis} as an
efficient method to calculate free energy differences. TeI is based on calculating for each system the 
$\ln  Z$ difference (where $Z$ is the partition function), between the temperature of interest and a high temperature using a Parallel Tempering procedure. Since at the high $T$ limit the two systems with the same degrees of freedom have the same partition function, the free energy difference can be calculated. In TeI in order to ensure the equation of the partition functions when $\beta\rightarrow 0$, we capped the potential terms - that is if $E$ was larger than $E_{\mathrm{cap}}$ it was set to $E_{\mathrm{cap}}$ (denoted in TeI by $E_{\mathrm{cutoff}}$). It is emphasized that the free energy difference calculated in TeI is between two different molecules while in relative free energy calculations the goal is to calculate free energy difference between the same molecule in two states (e.g solvated vs. unsolvated or bounded vs. unbounded) compared to the free energy difference of another molecule in these two states. While TeI has many advantages, that will also be apparent in this method, it cannot be directly applied to MD.  MD simulations at very high temperatures are impractical due to the very high velocities which will necessitate very small time steps for the integration of the equations of motion. In addition in TeI due to the simplicity of the energy models considered, effectively, all the energy terms are completely removed.

The presented method is based on TeI and is guided by the goal is to address the challenges previously mentioned - namely accuracy, simple implementation, robustness and high phase space overlap. First, in order to avoid reaching high temperatures (used in TeI), we use an additional variable $\lambda$ that will transform the system, keeping the temperature constant. Second, in order to enahnce phase space overlap we will keep some terms constant in the transformation.   

The method is based on first identifying in each of the compared systems (e.g molecule and environment) a common subsystem (e.g submolecule and the environment) and a different subsystem (e.g a submolecule). Then, each of the systems is transformed in each environment (e.g vacuum or water environment) by removing certain potential terms, into a system in which the partition functions of the common subsystem and the different subsystem can be decoupled. This \emph{decoupling} is not trivial as the atoms in one decoupled subsystem will still interact with atoms in another decoupled subsystem via potentials that relate between 2, 3 and 4 atoms. However, this will turn exact due to the fact that this is decoupling of the two integrals of the partition functions. This novel analysis is applicable to any potential function and to submolecules with coupled degrees of freedom, enabling to remove less terms in the transformation. Removing less terms in the transformation results in a smaller free energy difference which needs to be integrated over and higher phase space overlap (smaller statistical error), which are related to the efficiency of the calculation.
 Since each transformed system is simulated in two environments and the different subsystem can be treated as non-interacting system, the free energy associated with the different subsystem will analytically (exactly) cancel out in the thermodynamic cycle. Thus, instead of transforming between the compared systems to calculate the relative free energy difference, each system is transformed \emph{separately} into its replica with some energy terms relaxed (removed) and we avoid having ingredients (such as atoms and possibly force fields) of the two systems in the simulation. This ingredient is in fact a scheme for two separated \emph{topologies}. Transforming the systems separately has been suggested in recent works \cite{farhi2013general,liu2013lead} and is given here in a detailed mathematical description in the context of partition function decoupling. It is noted that the separate simulations are used here to calculate relative free energy (difference between two solvation/binding processes) and not only to calculate absolute solvation free energy (e.g Ref. \cite{schafer1999estimating}).  We demonstrate the decoupling analysis and the separate transformations in MD calculations of relative free energies which agree with experimental results. This decoupling analysis is also in agreement with our MD simulations in Ref. \cite{farhi2016calculation}. We then present a unified \emph{soft core} technique that will result in less steep integrated function. We prove analytically that the soft core technique ensures that the integrated function is monotonic. This statement regarding the monotonicity of the integrated function is novel and we then further extended to non-linear transformations. Since for monotonic functions the numerical integration error limit is known the integration result will be robust. In addition the monotonicity will enable simple selection of intermediates. We also show mathematically that in this soft core technique the numerically integrated function can be non-steep only when we transform the systems separately (novel result). Finally, we show that if the systems have rugged energy landscape, instead of using the sampling techniques such as \emph{H-REMD} in another $\lambda$ or $T$ dimension, we can use only one sampling dimension. 


The method is divided to its \emph{independent} ingredients. Namely, the decoupling analysis is applicable to the existing topologies. The topology ingredient can be used with the existing soft core schemes and the soft core ingredient can be used with the existing topologies. Each ingredient will be presented in a separate section with references to simulations that demonstrate it and to the state of the art corresponding ingredients. In Section \ref{section:decoupling} present an exact and complete \emph{decoupling} analysis. In Section \ref{section:two} we explain mathematically that the two systems can be simulated separately to give the free energy difference. This ingredient is related to \emph{topology} and can be called Two Topologies. In Section \ref{section:demonstration} we demonstrate the decoupling analysis and the separate transformations in MD simulations for a general force field. In Section  \ref{section:soft_core} we present a unified soft core technique \cite{FarhiThesis,buelens2012linear} and prove mathematically the monotonicity of the integrated function. In Appendix \ref{Appendix:improved_behavior} we show mathematically that when using this soft-core technique it is advantageous to transform the systems separately since the integrated function can be non-steep. 
In Section \ref{section:one_dimension} we explain how we can equilibrate the systems by using only one sampling dimension. In Section \ref{section:discussion} we summarize and discuss the method.

\section{Decoupling the partition functions}
\label{section:decoupling}
In this section we explain how by removing certain terms in the transformation, the partition function of the transformed system can be exactly decoupled into two partition functions. One partition function will be identical between the transformed systems at each environment and one will be of the different subsystem. This decoupling scheme is applicable to all topologies. The common subsystem is defined as an identical submolecule and the environment and the different subsystem 
is defined as the different submolecule between the compared systems. 
We will maximize the phase space overlap between the original and the transformed systems by removing as few terms as possible in the transformation (the phase space overlap is related to the number of intermediate systems needed in order to calculate the free energy difference). 



 The existing decoupling scheme for the hybrid topologies is based on the rigid rotor \emph{approximation} and the HJR technique \cite{boresch2003absolute}. In this scheme potential terms are removed in a transformation and then the system's partition function is decomposed into \emph{eight} partition functions - the partition function of the subsystem in common between the compared systems, the partition function(s) of  the different subsystem(s) (according to their definition) and a \emph{polymer-like} part which connects between them and decomposes into six partition functions \cite{boresch2003absolute,mugnai2012thermodynamic}. In case that the connecting part is not polymer-like, the potential terms which differentiate it from being polymer-like \emph{are removed} in the transformation. For example the methyl group is modeled with six bond angle terms and five of them \emph{need to be removed} in the transformation in order to conform to this requirement \cite{boresch2003absolute,mugnai2012thermodynamic}. In addition, the calculation of the six partition functions is considered and the potentials associated with them are \emph{required to be quadratic} \cite{boresch2003absolute,mugnai2012thermodynamic}. Thus, \emph{non-quadratic potential terms} such as dihedral terms (and possibly bond stretching and bond angle terms) \emph{are removed} in the transformation to enable the decoupling \cite{khavrutskii2010computing,mugnai2012thermodynamic}. This results in atoms which are bound to move on a sphere relatively to another atom instead of being properly located. When performing a larger transformation (removing more terms in the transformation), the computation time increases.


Here we present an \emph{exact} analysis in which we decouple the partition function of the transformed system into \emph{two} partition functions - the partition functions of the common and different subsystems. This analysis is applicable to \emph{any} potential function and thus the dihedral terms which are usually \emph{non-quadratic do not need to be removed} in the transformation. In addition, we show that when the different submolecule includes bond angle terms with coupled degrees of freedom (e.g a methyl group) it can also be exactly decoupled. Thus, \emph{coupled bond angle potentials do not need to be removed} in the transformation. Removing less terms in the transformation results in a smaller free energy difference and higher phase space overlap. Since the free energy difference and the phase space overlap are related to the the number intermediates and to the statistical error respectively, it is expected to reduce the computational power needed to perform the simulations. We note that in this analysis there is no need to consider calculation of integrals.

Molecular modeling includes covalent bond, bond angle, dihedral angle, improper dihedral, electrostatic and VDW potentials (see \cite{mayo1990dreiding,phillips2005scalable,hess2008gromacs,abraham2015gromacs} and Appendix A).
Covalent bond, bond angle and dihedral angle potential terms depend on the coordinates of two, three and four nearest covalently linked atoms respectively.
Electrostatic and VDW potentials relate between every atom pair in the system. Thus the energy terms can be separated into short range terms (covalent bond, bond angle, dihedral angle and improper dihedral angle) and long range terms (electrostatic and VDW). In the terminology of the field they are called bonded interactions and non bonded interactions respectively and we use these names in order to emphasize this difference between them.

To obtain the equilibrium constant, the standard Gibbs free energies are usually calculated. The standard state is the hypothetical state with the standard state concentration but exhibiting infinite-dilution behavior (the interactions between e.g the solute molecules are negligible). Hence, when we are interested in the properties of one substance \emph{a single copy of the molecule of interest can be simulated} either in vacuum or solvent environments.

We write the partition function of the system, which includes the molecule of interest and possibly the solvent molecules as follows:
$$Z=\int e^{-\beta H\left(\mathbf{\Omega}\right)}d\mathbf{\Omega}=\intop \prod_{i}^{l}e^{-\beta H\left(\mathbf{\Omega}\right)}d\mathbf{r}'_{i},$$
where $\mathbf{\Omega}$ is the coordinates vector of all the atoms in the systems, $\mathbf{r}'_{i}$ is the coordinate of atom $i$,  $\beta=\frac{1}{k_{B}T}$ and $k_B$ is Boltzmann constant. The integration is over all possible values of the vectors $\mathbf{r}'_{i}$.
We define the variables as follows:
\[
\mathbf{\Omega}=\left\{ \mathbf{r}'_{1},\mathbf{r}_{2},...,\mathbf{r}_{k},\mathbf{r}_{k+1},...,\mathbf{r}{}_{n},\mathbf{r}'{}_{n+1},...,\mathbf{r}'{}_{l}\right\} 
\]
where
 \begin{equation}
\mathbf{r}_{i} \equiv \mathbf{r}'_{i}-\mathbf{r}'_{i-1},
\end{equation}
which will be chosen as the position of atoms relative to covalently bounded atoms (bold letters denote vectors).  $\mathbf{r}'_k$ represents the position of the last atom that is common between the compared systems and $\mathbf{r}'_{k+1}$ represents the position of the first atom in the different submolecule. $n$ denoted the index of the last atom in the molecule and $\left\{ \mathbf{r}'_{n+1},...,\mathbf{r}'{}_{l}\right\} \equiv \mathbf{\Omega}_{\mathrm{env}}$ denote (if necessary) the solvent molecules. The partition function can now be written as follows:
$$Z=\intop e^{-\beta H\left(\mathbf{\Omega}\right)}d\mathbf{r}'_{1}\prod_{i=2}^{k}d\mathbf{r}_i\prod_{j=k+1}^{n}d\mathbf{r}_{j}\prod_{m=n+1}^{l}d\mathbf{r}'_{m}
 ,$$ 
 Integration over these degrees of freedom will of course give the same result.

\begin{figure}[h]
 \centering
\includegraphics[width=8cm]{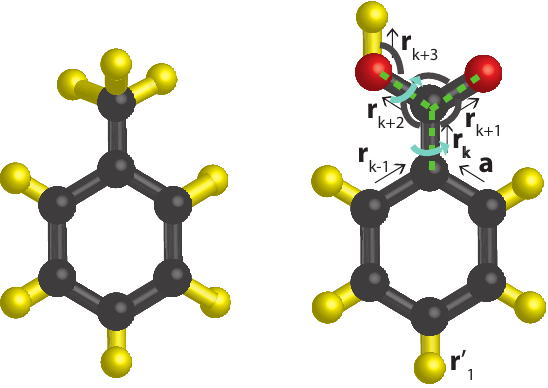}
 \caption{An example of the new coordinates of the atoms in Benzoic Acid in comparison to Toluene. The short range interactions that include the different atoms are plotted - the bond angle, dihedral and improper dihedral terms  are marked by arcs, arcs with arrows and three intersecting dashed lines respectively.}
 \label{fig:BenzoicAcidCoordinates}
\end{figure}
To illustrate the technique we compare the molecules Benzoic Acid and Toluene which include an aromatic ring and few different atoms. The method presented can be applied to molecules in which there is one separation point between the common and different submolecules (the different atoms do not form a loop that starts at one atom and ends in another atom in the molecule). The numbers of atoms of the two compared molecules can be different since the free energies associated with the different subsystems will cancel out in the Thermodynamic Cycle (Section \ref{section:two}, Fig. \ref{fig:summary}). The decoupling analysis will be demonstrated in a relative free energy calculation with two other molecules in Section \ref{section:demonstration}.

In Fig. \ref{fig:BenzoicAcidCoordinates} an example of the new coordinates of the atoms in the molecule Benzoic Acid in its comparison to Toluene is presented. The vector $\mathbf{a}$ denotes the relative coordinate of the top atom in the ring, and it is used since $\mathbf{r}_{k-1}$ represents the relative position of this atom with respect to another atom. We will use the notations on Benzoic Acid in the figure in the next explanations and the following analysis is applicable also to Toluene.

We will now turn to explain how the system's partition function can be separated into two partition functions identically - the partition function of the common submolecule and the environment and the partition function of the different submolecule.

We first define the coordinates of the atoms of the common submolecule as $\left(\mathbf{r}'_{1},\mathbf{r}_{2},...,\mathbf{r}_{k}\right)$ and the coordinates of the atoms of the different submolecule as $\left(\mathbf{r}_{k+1},\mathbf{r}_{k+2}, \mathbf{r}_{k+3}\right)$.
In the following analysis it is assumed that in the transformed state the interactions of the common submolecule with itself and the environment are kept constant. We will also assume that there are no improper dihedral and long range terms that couple atoms from the two subsystems.  In Section \ref{section:two} it will be explained how these terms are removed in the transformation.

When a system can be separated into two groups of particles that are not interacting with each other it is well known that its partition function can be separated into the partition functions of the two groups of particles and be written as their multiplication. Thus, it is clear that in the transformed state the existence of interactions between the atoms in the different submolecule will not prevent us from separating the partition function into two partition functions. Hence, terms that involve only atoms of the different submolecule can remain constant in the transformation. This also includes the long range terms between the atoms in the different submolecule (in agreement with Ref. \cite{mugnai2012thermodynamic}). It is worth noting that transforming the molecule into a molecule with a different charge in the context of MD can result in free energy change due to artifacts that originate from the periodicity of the system \cite{hunenberger1999ewald}.

We now turn to explain that three types of short range terms that involve atoms from the different submolecule, including ones that couple the different and common submolecules, can remain constant in the transformation and will enable us to decouple the partition function into the two. 
In standard molecular modeling there are the following covalent bond terms that depend on the positions of the different atoms: $V_{c}\left(\mathbf{r}_{k+1}\right),V_{c}\left(\mathbf{r}_{k+2}\right)$ and $V_{c}\left(\mathbf{r}_{k+3}\right)$. The bond angle terms that depend on the position of the different atoms are: $V_{b}\left(\mathbf{r}_{k},\mathbf{r}_{k+1}\right),V_{b}\left(\mathbf{r}_{k},\mathbf{r}_{k+2}\right),V_{b}\left(\mathbf{r}_{k+1},\mathbf{r}_{k+2}\right)$  and $V_{b}\left(\mathbf{r}_{k+2},\mathbf{r}_{k+3}\right)$ . Only one of the following dihedral terms $V_{d}\left(\mathbf{r}_{k-1},\mathbf{r}_{k},\mathbf{r}_{k+1}\right) , V_{d}\left(\mathbf{r}_{k-1},\mathbf{r}_{k},\mathbf{r}_{k+2}\right) , V_{d}\left(\mathbf{a},\mathbf{r}_{k},\mathbf{r}_{k+1}\right)$  and $V_{d}\left(\mathbf{a},\mathbf{r}_{k},\mathbf{r}_{k+2}\right)$  is usually used since they model the rotation of the different submolecule in which the bond angles are kept constant . We will assume for the following explanation that the dihedral term $V_{d}\left(\mathbf{r}_{k-1},\mathbf{r}_{k},\mathbf{r}_{k+2}\right)$ is used. The dihedral term $V_{d}\left(\mathbf{r}_{k},\mathbf{r}_{k+2},\mathbf{r}_{k+3}\right)$ usually also exists. In addition the improper dihedral term $V_\mathrm{i\,d}\left(\mathbf{r}_{k},\mathbf{r}_{k+1},\mathbf{r}_{k+2}\right)$ usually exists and will be removed in the transformation (see Fig. \ref{fig:BenzoicAcidCoordinates}).


If we define $$\hat{z}\equiv-\hat{r_k},$$ the bond angle terms that depend on $\mathbf{r}_k$ can depend instead on  $\hat{z}$.

\begin{figure}[h]
 \centering
\includegraphics[width=8cm]{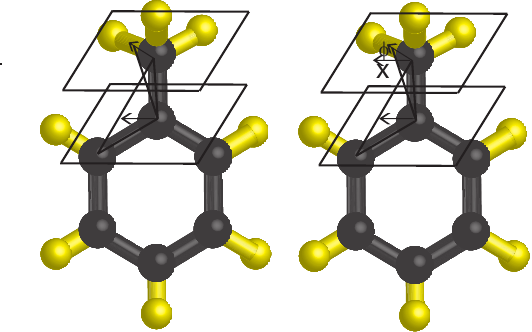}
 \caption{An illustration that shows the correspondence between the dihedral angle and $\phi$ angle. The triangles are in the planes between which the dihedral angle is measured. The rectangles are perpendicular to the line of intersection of these planes. The vectors are both in these planes and in the planes which are perpendicular to the intersection line. In the molecule on the right, the bottom vector is placed next to the one on top in order to show the correspondence between the angles.}
 \label{fig:explanation_angle}
\end{figure}

The dihedral potential term depends on the angle between two planes (see Appendix A for details) which can be defined as the angle between the vectors in these planes that are perpendicular to the intersection line of these planes. We notice that the dihedral angle $\phi\left(\mathbf{r}_{k-1},\mathbf{r}_{k},\mathbf{r}_{k+2}\right)$ is equal to $\phi$ angle in spherical coordinates defined with respect to $\hat{z}$ and
$$\hat{x}\equiv -\frac{\mathbf{r}_{k-1}-\left(\mathbf{r}_{k}\cdot\mathbf{r}_{k-1}\right)\hat{r}_{k}}{\left|\mathbf{r}_{k-1}-\left(\mathbf{r}_{k}\cdot\mathbf{r}_{k-1}\right)\hat{r}_{k}\right|}$$
(see Fig. \ref{fig:explanation_angle}).
Hence the corresponding dihedral term depends on $\hat{x},\hat{z}$ and $\mathbf{r}_{k+2}$.

We define ${\mathbf{\Omega}}_{\mathrm{{com}}}=\left(\mathbf{r}'_{1},\mathbf{r}_{2},...,\mathbf{r}_{k},{\mathbf{\Omega}}_{\mathrm{env}}\right)$ and $H_{\mathrm{com}}$ as all the potential terms that depend on the coordinates of the atoms in the common subsystem $\mathbf{\Omega}_{\mathrm{com}}$ . We write the partition function as follows:
\begin{gather}
Z=\int e^{-\beta H_{\mathrm{com}}\left(\mathbf{\Omega}_{\mathrm{com}}\right)}d\mathbf{\Omega}_{\mathrm{com}}\int\prod_{j=k+1}^{k+3}e^{-\beta V_{c}\left(\mathbf{r}_{j}\right)}\times \nonumber \\
e^{-\beta\left[V_{b}\left(\hat{z},\mathbf{r}_{k+1}\right)+V_{b}\left(\hat{z},\mathbf{r}_{k+2}\right)+V_{b}\left(\mathbf{r}_{k+1},\mathbf{r}_{k+2}\right)+V_{b}\left(\mathbf{r}_{k+2},\mathbf{r}_{k+3}\right)\right]}\times \nonumber \\
e^{-\beta\left[V_{d}\left(\hat{x},\hat{z},\mathbf{r}_{k+2}\right)+V_{d}\left(\hat{z},\mathbf{r}_{k+2},\mathbf{r}_{k+3}\right)\right]}d\mathbf{r}_{j}.
 \end{gather}

\begin{figure}[h]
 \centering
\includegraphics[width=8cm]{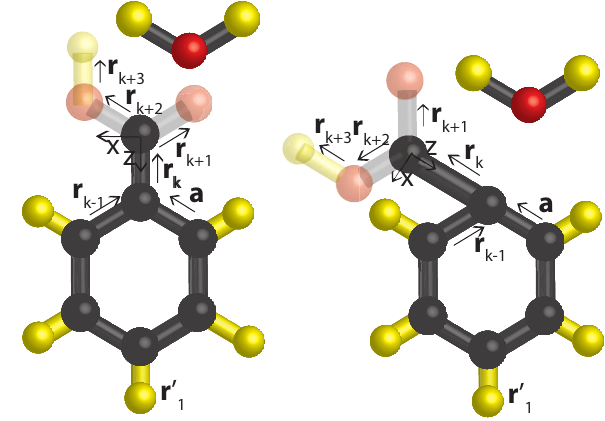}
 \caption{Two sets of coordinates of the common subsystem. The transparent atoms represent atoms whose long range and improper dihedral interactions with the common subsystem are removed. The atoms of the common subsystem, including the water molecules in the case of explicit solvent can enter the volume of the different atoms since the long range interactions between them are relaxed. $Z_{\mathrm{dif}}$  is the same in both cases.}
 \label{fig:explanation_integral}
\end{figure}

We now notice that given a set of coordinates of the common submolecule, the only information used in the integration over $\Omega_{\mathrm{dif}}=\left\{ \ensuremath{\mathbf{r}_{k+1},\mathbf{r}_{k+2}},\mathbf{r}_{k+3}\right\}$ 
  is the orientation of $x$  and $z$ axes. This information does not affect the integration result since the different submolecule does not have a preferred direction. In other words, for any set of coordinates of the common subsystem the integration result over $\Omega_{\mathrm{dif}}$ is the same (see Fig. \ref{fig:explanation_integral}). We can thus write:
\begin{equation}
Z=\int e^{-\beta H_{\mathrm{com}}\left(\mathbf{\Omega}_{\mathrm{com\,}}\right)}d\mathbf{\Omega}_{\mathrm{com}}Z_{\mathrm{dif}}=Z_{\mathrm{com}}Z_{\mathrm{dif}},
\end{equation}
where $Z_{\mathrm{com}}$ and  $Z_{\mathrm{dif}}$  denote the partition functions of the common and different subsystems respectively. Note that  $Z_{\mathrm{dif}}$ is a constant and does not affect the integration over $\mathbf{\Omega}_{\mathrm{com}}$. See detailed proof in Appendix \ref{appendix:detailed_proof}.

 Hence, the bond stretching, bond angle and dihedral angle energy terms, effectively, do not couple the partition functions. This is despite the fact that some of these terms involve atoms from the two submolecules.

 This decoupling of the partition functions does not depend on the potential function but only on the variables it depends on. Moreover, the partition function of submolecules with coupled degrees of freedom such as the three bond angle terms associated with the vectors $\mathbf{r}_k,\mathbf{r}_{k+1}$ and $\mathbf{r}_{k+2}$ decouple exactly. 

In case there will be a dihedral potential term that depends on the vector $\mathbf{r}_{k-1}$ (e.g the dihedral term defined by $\mathbf{r}_{k-1},\mathbf{r}_{k}$ and $\mathbf{r}_{k+2}$) and a dihedral term that depends on $\mathbf{a}$ (e.g the dihedral term defined by $\mathbf{a},\mathbf{r}_{k}$ and $\mathbf{r}_{k+1}$), there will be dependence between the partition functions since the coordinates of the atoms in the common submolecule will determine $x,z$ axes but also another vector in the $xy$ plane which is related to the configuration of the common submolecule. However, usually only one dihedral angle energy term is used to relate between such subsystems and this explanation is given for generality.

 A similar analysis applied to the improper dihedral terms that relate between atoms in the two submolecules shows that the partition function of the different subsystem \emph{does} depend on the coordinates of the atoms in the common subsystem. Thus, it is suggested that modeling the planarity of the molecule with dihedral angle terms rather than improper dihedral terms (optional in Gromacs manual \cite{abraham2015gromacs}) will result in less removed terms in the transformation. 

\begin{figure}[h!]
 \centering
\includegraphics[width=8cm]{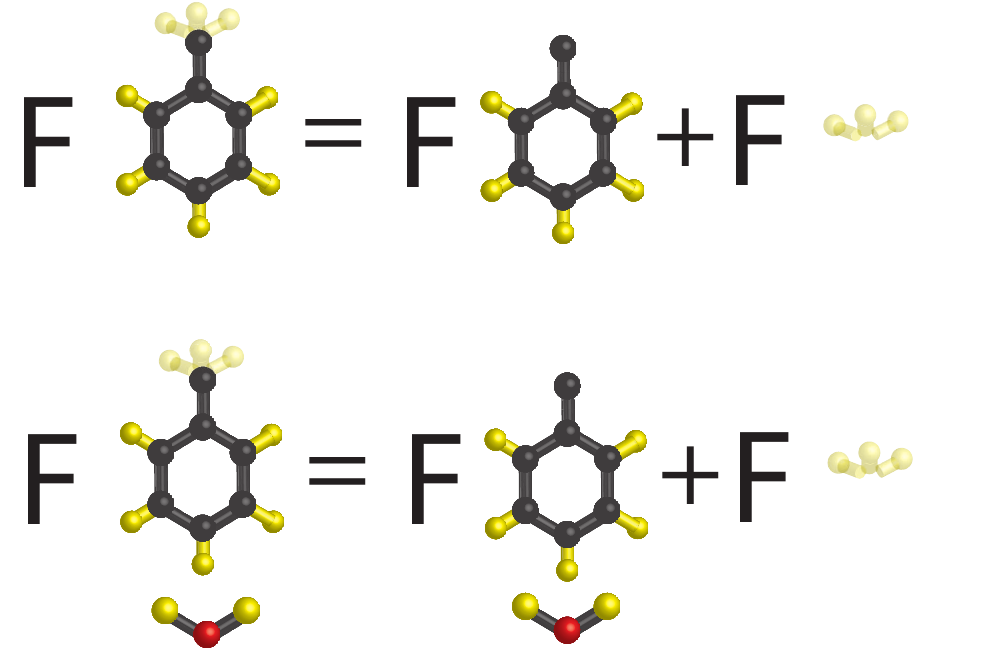}
 \caption{A scheme of the free energy of the transformed Toluene in two environments (F denotes free energy)}
 \label{fig:TolueneTransformedFreeEnergy}
\end{figure}

We can write in terms of the partition functions:
\begin{equation}
Z\left(\mathbf{r}'_{1} ,\mathbf{r}{}_{2},..,\mathbf{r}_{n},\mathbf{\Omega}_{\mathrm{env}}\right)\rightarrow
\end{equation}
$$ Z_{\mathrm{common\, int}}\left(\mathbf{r}'_{1},\mathbf{r{}}_{2},..,\mathbf{r}_{k},\mathbf{\Omega}_{\mathrm{env}}\right)Z_{\mathrm{diff\, non\, int}}\left(\mathbf{r}_{k+1,...,}\mathbf{r}_{n}\right),$$
where $Z_{\mathrm{common\, int}}$ denotes the partition function of the common subsystem in which the common submolecule interacts with the environment, $\mathbf{\Omega}_{\mathrm{env}}$ denotes the coordinates of the molecules of the environment, $Z_{\mathrm{diff\, non\, int}}$ denotes the partition function
of the different submolecule that does not interact with
the environment and the arrow symbolizes the transformation in which energy terms are relaxed.
We define $A$ and $B$ to be the compared systems in a certain environment and $A'$ and $B'$ as their transformed replicas (the systems without the terms that couple the partition functions as previously explained). 
It can thus be written  (see Fig. \ref{fig:TolueneTransformedFreeEnergy}):
\begin{equation}
F_{A'}=F_{A'_{\mathrm{common\, int}}}+F_{A'_{\mathrm{diff\, non\, int}}},\nonumber
\end{equation}
\begin{equation}
F_{B'}=F_{B'_{\mathrm{common\, int}}}+F_{B'_{\mathrm{diff\, non\, int}}}.
\end{equation}
It is noted that in fact the dihedral and bond angle terms that include atoms from the common and the different submolecules are associated with $Z_{\mathrm{diff\, non\, int}}\left(\mathbf{r}_{k+1,...,}\mathbf{r}_{n}\right)$. The orientations of the $\mathbf{r}_k,\mathbf{r}_{k+1}$ vectors that are in the common submolecule, are also effectively associated with the different submolecule in the form of arbitrary orthogonal $x$ and $z$ axes for the separate theoretical calculation of free energy.

In the case of totally different molecules it can thus be written:
\[
Z\rightarrow Z_{\mathrm{diff\, non\, int}}.
\]
In the terminology of the field removing the long range energy terms is called \emph{decoupling} and the decoupled atoms are called \emph{dummy}. In this context we can call the different subsystem \emph{dummy subsystem} since we can keep the internal  energy terms in the different subsystem constant in the transformation. In addition, in some cases in binding the long range terms between the different subsystem and the common environment (water) may be kept constant and may have a small effect on the free energy. However, this assumes that the water molecules that are interacting with the different subsystem are weakly interacting with the common submolecule and is therefore not recommended in the general case.

In standard molecular modeling the force field parameters of the common subsystems are identical, and hence the common subsystems are identical. 
In Fig. \ref{fig:TolueneTransformedFreeEnergy} a scheme of the free energies of the transformed replicas of Toluene in vacuum and water environments are presented. It can be seen that in both environments the free energy of the transformed replica can be decomposed into the free energy of the common and the different subsystems. The free energy of the different subsystem is equal in the two environments since it effectively does not interact with the rest of the system.

To summarize the following terms can remain constant in the transformation and yield an exact calculation, \emph{independently} of the potential function:
\begin{enumerate}
\item Bond stretching and bond angle terms.
\item Dihedral energy terms, as long as terms that involve atoms from the two submolecules are associated with two covalent bonds in the common submolecule (e.g $\mathbf{r}_k$ and $\mathbf{r}_{k-1}$).\
\item Potential terms that involve only atoms of the different subsystem.
\end{enumerate}

We now denote the Hamiltonian with all the terms that are removed in the transformation by $H_r$. This Hamiltonian includes the VDW and electrostatic interactions of the different atoms with the rest of the system and improper dihedral terms (in case they are defined in the usual manner) that relate between atoms from the different and common submolecules. We denote by $H_c$ the other terms in the system. These definitions will be used in the next section.

\subsection*{Verification of the analysis with MD simulations}

The analysis presented in this section is in agreement with the MD simulation results in Ref. \cite{farhi2016calculation} in which free energies associated with \emph{non-quadratic} potentials for a methanethiol molecule were calculated. The free energies associated with bond angle potential, methyl group with coupled bond angle potentials and dihedral angle potential in the transformed state were the same in vacuum and water environments. This means that the partition functions which include these terms decouple. This is in agreement with the analysis presented in this section according to which such terms do not have to be removed in the transformation in order for the partition functions to decouple.

\section{Two Separate Simulations}
\label{section:two}
In relative free energy calculations the difference between solvation/ binding processes of two similar molecules is calculated. Our goal here is to calculate the relative free energy by transforming each system \emph{separately} without having ingredients  such as atoms and force fields from the two compared molecules in the simulation (see Fig. \ref{fig:HybridTransformation}). These separate simulations can facilitate automation as it eliminates the need for human intervention in setting up the simulation.  Transforming the systems separately has been suggested in recent works \cite{farhi2013general,liu2013lead} and is given here in a detailed mathematical description in the context of partition function decoupling.

\begin{figure}[h!]
 \centering
\includegraphics[width=8cm]{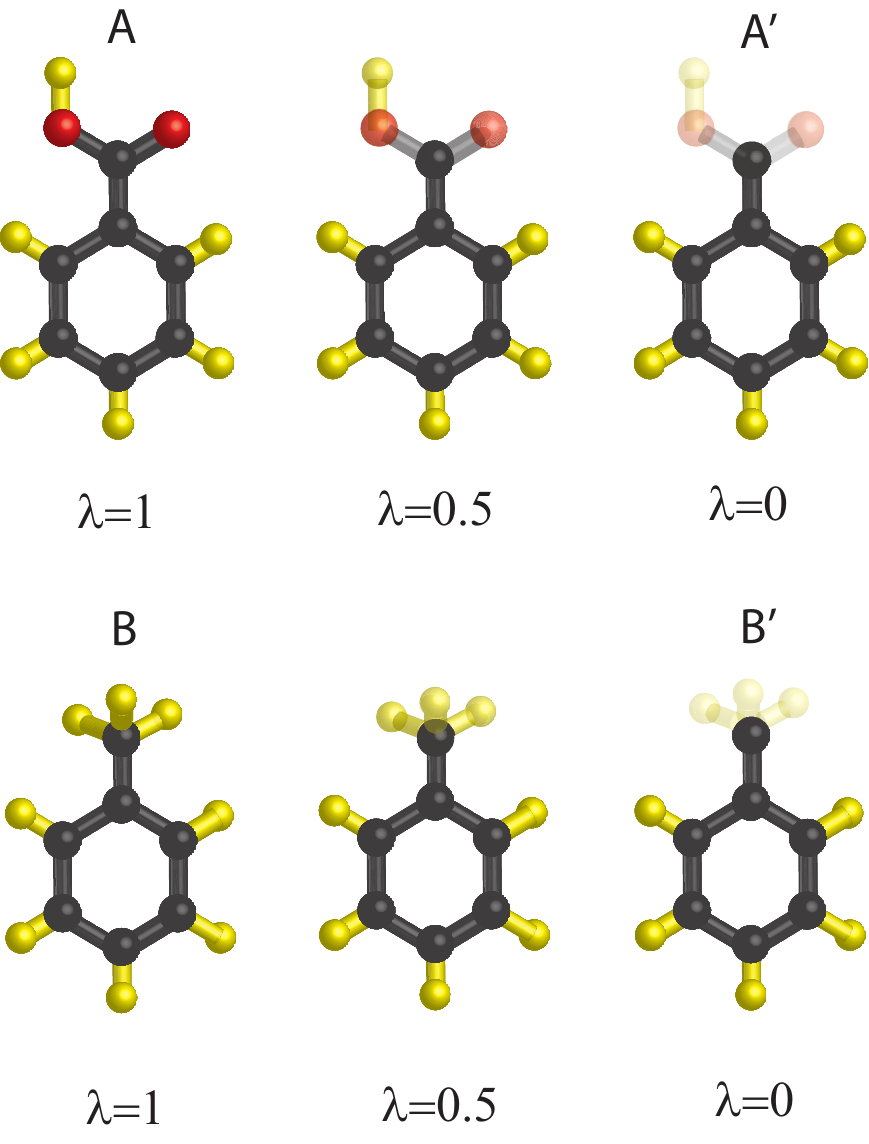}
 \caption{A scheme of the transformations in the novel method (two systems). $A$ and $B$ are the compared systems and $A'$ and $B'$ are the transformed replicas - the original systems with the terms denoted by $H_r$ removed.  The transparent submolecule is the decoupled submolecule as previously explained.}
 \label{fig:SeparateTransformation}
\end{figure}

The idea in this section is to first identify in each of the compared systems a common subsystem and a different subsystem. Then, to transform each of the systems in each environment (e.g vacuum and water for solvation) by relaxing the Hamiltonian $H_r$, into a system in which the partition functions of the common subsystem and the different subsystem can be decoupled. In this section we assume that the potential terms that diverge at $r\rightarrow0$, such as the VDW and electrostatic interactions, when we are close to the transformed state, do not play a role. In the next section we will justify this assumption. In Fig. \ref{fig:SeparateTransformation} a scheme of the two separate systems suggested in this section is presented. 
We now explain how the free energy difference between the original systems and their transformed replicas can be calculated in the context of TI. We emphasize that it can also be calculated with other free energy calculation methods (FEP, BAR etc.).
The $\lambda$ dependent Hamiltonians can be written as follows (please note that each equation includes a definition for the system $A$ and a definition for the system $B$):
\begin{equation}
\label{eq:second_transformation}
H_{A/B}\left(\lambda\right)=\lambda
H_{_{{A_{r}/B_{r}}}}+H_{{}{A_c/B_c}},
\end{equation}
where the Hamiltonian with the terms that are removed in the transformation and the Hamiltonian including all the other terms are denoted by $H_r$ and $H_c$ respectively as previously explained.

The free energy between each system and its transformed replica can be written as follows:
\begin{equation*}
\Delta F_{A/B\rightarrow A'/B'}=
\end{equation*}
\begin{equation*}
k_{B}T\left[\ln Z_{A/B}\left(\beta,\lambda=1\right)-\ln Z_{A/B}\left(\beta,\lambda=0\right)\right]=
\end{equation*}
\begin{equation}
\label{eq:integration}
k_{B}T\int_{0}^{1}\frac{d\ln Z_{A/B}\left(\beta,\lambda\right)}{d\lambda}d\lambda=-\int_{0}^{1}\left\langle H_{_{{A_{r}/B_{r}}}}\right\rangle d\lambda.
\end{equation}\
Each system is simulated at a set of $\lambda$s in the range $[0,1]$ that interpolates between the original and transformed systems. The average energy $ \left\langle H_{_{{A_{r}/B_{r}}}}\right\rangle$ is calculated in the simulation at each $\lambda$ value. Thus, we can numerically integrate and get the free energy difference between the original and transformed systems. It is emphasized that the free energy difference calculated here is between $A$ and $A'$ or between $B$ and $B'$ as opposed to the standard calculation in which the free energy is calculated between $A$ and $B$ as in Eq. (\ref{eq:first_transformation}).


We now denote the first and second environments by $1,2$ respectively. Examples for two environments are vacuum and water for solvation free energy and water and protein+water for binding free energy. We denote the compared systems in each environment by $A_1,B_1,A_2,B_2$ and their transformed replicas by $A_{1}^{'},B_{1}^{'},A_{2}^{'},B_{2}^{'}$. According to the explanations in the previous section, taking into account that the partition functions of the different subsystems are not coupled to the environment, we write:
\begin{equation}
A_{\mathrm{1}}\rightarrow A_{1}^{'},B_{1}\rightarrow B_{1}^{'} \,A_{2}\rightarrow A_{2}^{'},B_{2}\rightarrow B_{2}^{'},
\end{equation}
\begin{equation}
Z_{A_{1}^{'}}=Z_{\mathrm{1_{identical}}}Z_{A_{\mathrm{different}}^{'}},\, Z_{B_{1}^{'}}=Z_{\mathrm{1_{identical}}}Z_{B_{\mathrm{different}}^{'}}, 
\end{equation}
\begin{equation}
Z_{A_{2}^{'}}=Z_{\mathrm{2_{identical}}}Z_{A_{\mathrm{different}}^{'}},\, Z_{B_{2}^{'}}=Z_{\mathrm{2_{identical}}}Z_{B_{\mathrm{different}}^{'}}.
\end{equation}
It can be seen that the partition functions of the transformed replicas are decomposed to sub partition functions which are equal to sub partition functions of other transformed replicas (having the same names).
\begin{figure}[h]
 \centering{}
\includegraphics[height=6cm, width=8cm]{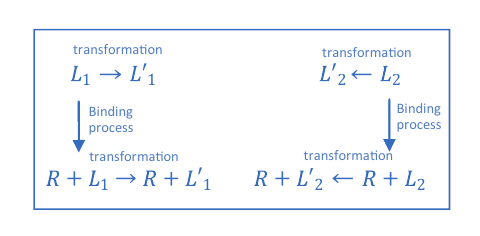}
 \caption{A scheme of the free energy differences in the novel method}
 \label{fig:bindingh}
\end{figure}
In Fig. \ref{fig:bindingh} a scheme of the free energies in the novel method is presented. $R+L_1,R+L_1$ denote the ligands bound to the receptor. $L_1,L_2$ denote the unbound ligands in an environment that does not include the receptor. The unbound receptor does not have to be simulated since it is the same in both $L_1$ and $L_2$ end states. The ligand $L_1$/ $L_2$ is transformed into its replica $L_1'$/ $L_2'$ with the $H_r$ terms completely relaxed. The end states $L_1,L_2,L_1',L_2',R+L_1,R+L_2,R+L_1',R+L_2'$ correspond to the end states $A_1,B_1,A_{1}^{'},B_{1}^{'},A_2,B_2,A_{2}^{'},B_{2}^{'}$ respectively.

\begin{figure}[h!]
 \centering
\includegraphics[width=8cm]{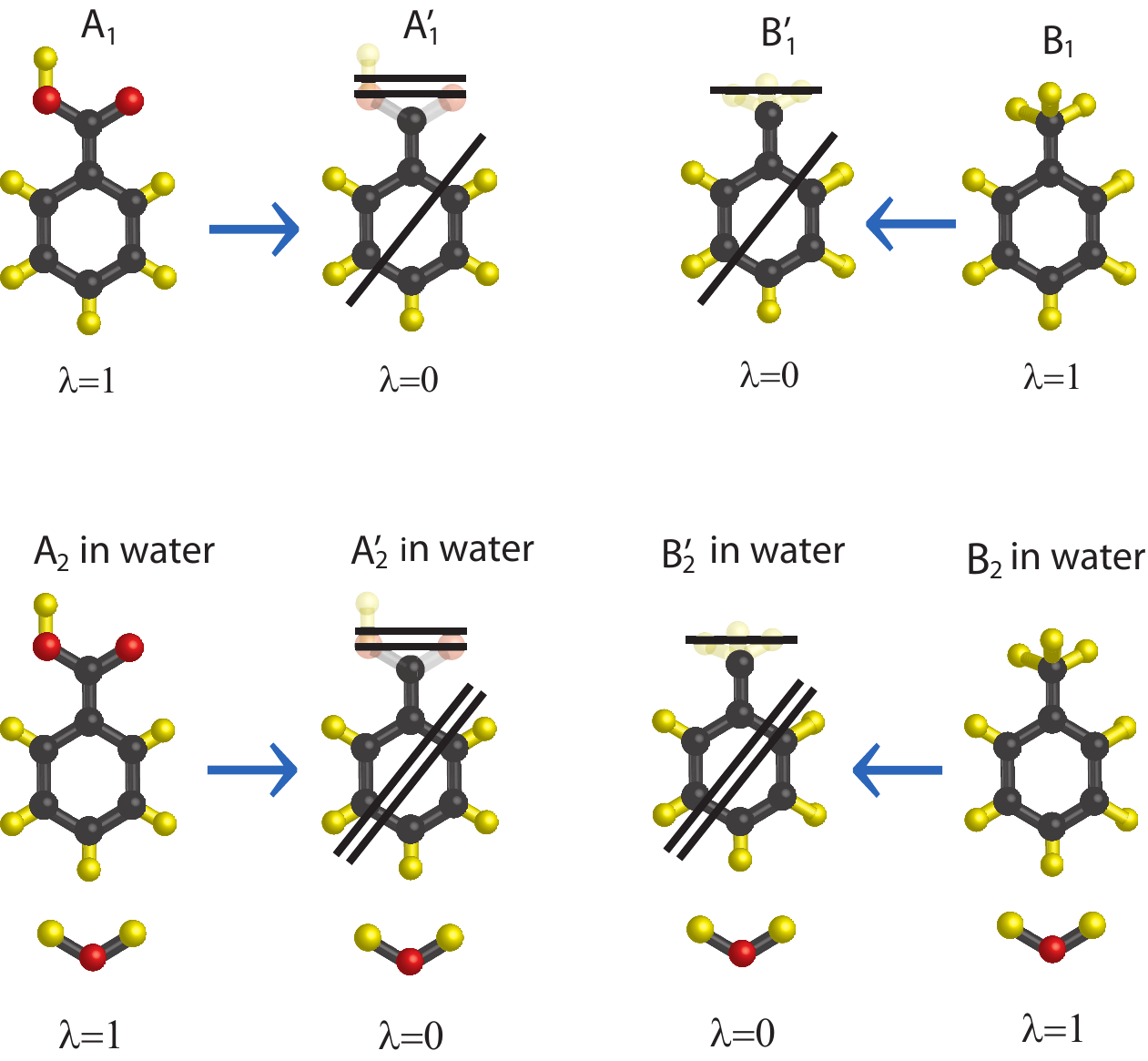}
 \caption{A summary scheme of the transformations and free energy cancellations (context of solvation)}
 \label{fig:summary}
\end{figure}

The free energy difference between  $L_1'$ and  $L_2'$ is equal to the free energy difference between $R+L_1'$ and $R+L_2'$ ($\Delta F_{L'_1\rightarrow L'_2}=\Delta F_{R+L_1'\rightarrow R+ L'_2}$), since the free energy of the common subsystems in each environment are equal and the free energy of the different subsystems is the same in both environments.
This can be written explicitly as follows:
\begin{equation*}
\Delta F_{L'_{1}\rightarrow L'_{2}}=
\end{equation*}
\begin{equation}
-k_{B}T\left(\ln Z_{\mathrm{w\, id}}+\ln Z_{L_{2\,\mathrm{dif}}^{'}}-\ln Z_{\mathrm{w\, id}}-\ln Z_{L_{1\,\mathrm{dif}}^{'}}\right)= \\\nonumber
\end{equation}
\begin{equation}
-k_{B}T\left(\ln Z_{L_{2\,\mathrm{dif}}^{'}}-\ln Z_{L_{1\,\mathrm{dif}}^{'}}\right)= \\\nonumber
\end{equation}
\begin{equation}
-k_{B}T\left(\ln Z_{\mathrm{wp\, id}}+\ln Z_{L_{2\,\mathrm{dif}}^{'}}-\ln Z_{\mathrm{wp\, id}}-\ln Z_{L_{1\,\mathrm{diff}}^{'}}\right)= \\\nonumber
\end{equation}
\begin{equation}
\Delta F_{R+L_{1}'\rightarrow R+L'_{2}}, \\ 
\end{equation}
%
where the subscripts ``w'' and ``wp'' stand for water environment and water protein environment respectively. The subscripts ``id'' and ``dif'' denote the identical and different subsystems respectively.
The relative free energy can now be written as follows:
\begin{equation*}
\Delta F_{A_{\mathrm{solvation/binding}}\rightarrow B_{\mathrm{solvation/binding}}}= 
\end{equation*}
\begin{equation*}
\Delta F_{L_{1}\rightarrow R+L_{1}}-\Delta F_{L_{2}\rightarrow R+L_{2}}=
\end{equation*}
\begin{equation*}
\Delta F_{L_{1}\rightarrow L'_{1}}+\Delta F_{L'_{1}\rightarrow L'_{2}}-\Delta F_{L_{2}\rightarrow L'_{2}}
\end{equation*}
\begin{equation*}
-\left(\Delta F_{R+L_{1}\rightarrow R+L'_{1}}+F_{R+L_{1}'\rightarrow R+L'_{2}}-\Delta F_{R+L_{2}\rightarrow R+L'_{2}}\right)= 
\end{equation*}
\begin{equation*}
\Delta F_{L_{1}\rightarrow L'_{1}}-\Delta F_{L_{2}\rightarrow L'_{2}}
\end{equation*}
\begin{equation*}
-\left(\Delta F_{R+L_{1}\rightarrow R+L'_{1}}-\Delta F_{R+L_{2}\rightarrow R+L'_{2}}\right)=
\end{equation*}
\begin{equation*}
\int_{0}^{1}\left\langle H_{B_{r}}\right\rangle -\int_{0}^{1}\left\langle H_{A_{r}}\right\rangle 
\end{equation*}
\begin{equation}
+\int_{0}^{1}\left\langle H_{A_{\mathrm{solvated/bounded}_{r}}}\right\rangle -\int_{0}^{1}\left\langle H_{B_{\mathrm{solvated/bounded}_{r}}}\right\rangle . 
\label{eq:relative_free_energies}
\end{equation}
Thus we perform in each environment  (in vacuum and in solvent for solvation and in solvent and in solvent-protein environment for binding) one transformation of each of the compared molecules.
That is, each molecule is simulated at a set of $\lambda$s and the free energy between the original and transformed systems is calculated. These free energy differences will allow us to get the relative free energy difference.
Fig. \ref{fig:summary} is a summary scheme of the transformations and free energy cancellations. As previously mentioned the cancellations of the free energies are exact. It is emphasized that (in all topologies) there is no restriction on the number of atoms of the compared molecules since these factors cancel out in the Thermodynamic Cycle. 
In each of the simulations we have only the atoms and force fields of the original molecule. Thus human intervention in order to integrate between the systems and their force fields and to disable the interaction between the systems is not required. Here we achieve a considerable simplification over the simulation in the hybrid system setup, and a simplification and significantly better phase space overlap as compared with the dual topologies setup since less terms are removed.   

It is noted that this ingredient is different from the single reference comparison schemes \cite{khavrutskii2010computing} in which all the molecules are compared with one reference molecule. This is since the single reference comparison is defined for a specific group of molecules and the reference system has to be given as an input and is not generated from the compared molecules.  Using the two original systems for the comparison allows  generality and flexibility - can be used for any group of two or more molecules. The use of the molecules themselves as a reference ensures maximal phase space overlap. The separation of the system into the common and different submolecules ensures that we maintain the largest common subsystem untransformed. This is as opposed to previous works in which the reference molecule is not the largest common subsystem (e.g Ref.   \cite{khavrutskii2010computing}).


\section{Demonstrations of the decoupling analysis and the separate transformations} 
\label{section:demonstration}

We now demonstrate the decoupling analysis and the separate simulations in calculating the free energy difference between solvation of para-Cresol and para-Cholorophenol. In the force field we used all the potential terms that depend on the spherical variables \emph{are non-quadratic} 
\[
V_{c}=\frac{1}{2}k_{c}\left(r^{2}-r_{0}^{2}\right)^{2},\, V_{b}=\frac{1}{2}k_{\theta}\left(\cos\theta-\cos\theta_{0}\right)^{2},\, 
\]
$$V_{d}\left(\phi_{ijkl}\right)=k_{\phi}\left(1+\cos\left(n\phi-\phi_{s}\right)\right).$$
 Therefore, the calculation presented here can be performed only when using the decoupling analysis presented in Section \ref{section:decoupling}.

In standard force fields the parameters of the common subsystems are usually identical between the compared systems.
We consider here the general case in which the molecules are individually parametrized by Quantum Mechanics computations \cite{malde2011automated}. Thus the common submolecules do not necessarily have the same terms.


\begin{figure}[h]
\includegraphics[width=8cm]{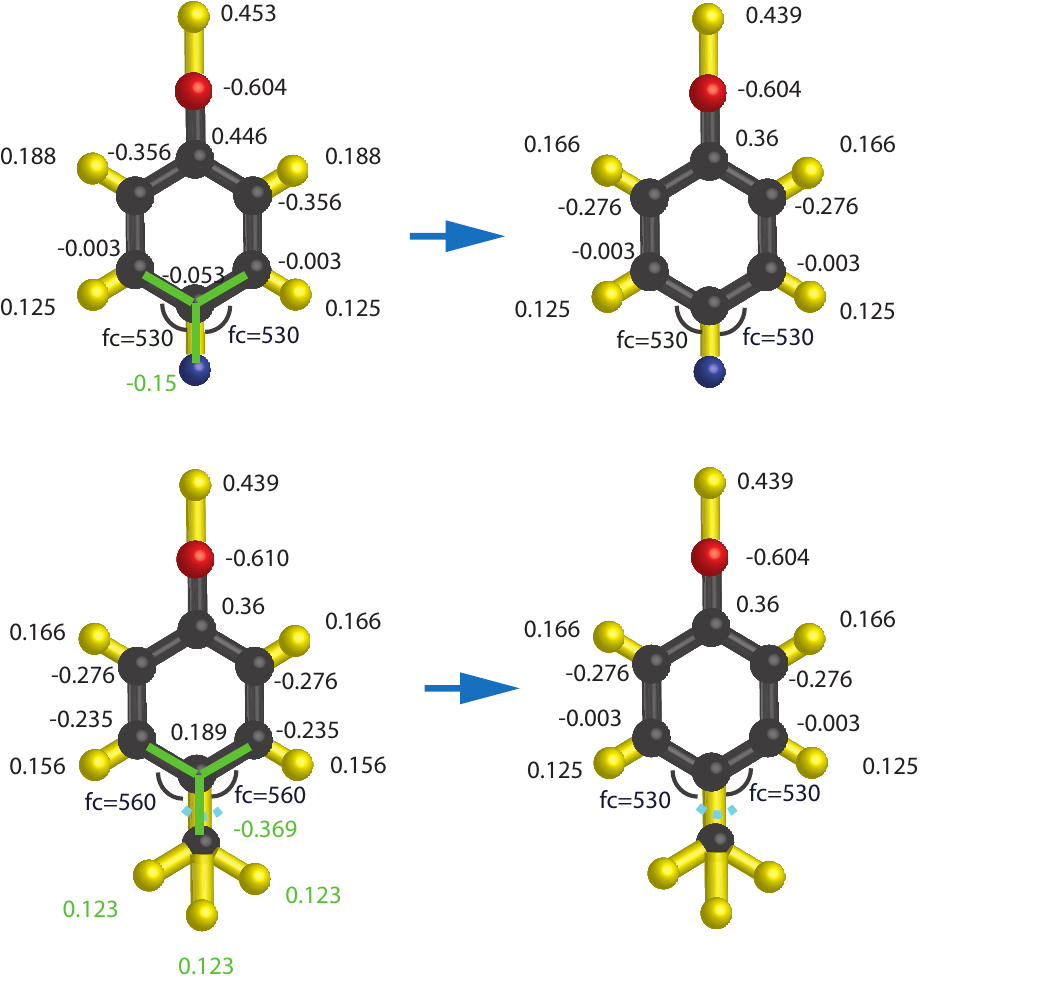}
\caption{Illustration of the transformations. The partial charges written in black and green belong to the common and different subsystems respectively. The light blue dashed line denotes the dihedral term. The three connected green lines represent the improper dihedral term. In case there is no term its value is zero.}

\end{figure}

%
%
%
%
%

Figure 10 is an illustration of the transformations into systems in which the common subsystems are identical and the partition functions of the different and common subsystem decouple. 
 The VDW and electrostatic potentials of the different atoms as well as the improper dihedral term (three connected green lines) couple between the two subsystems and are therefore relaxed in the transformation.
The bond angle terms marked on the figure have the same $\theta_0$ and different $k_b$
(corresponding to $f_{c}$), which after the transformations are the same.
 The VDW terms are not presented since they are the same for the same atoms.

 Similar transformations apply for a group of 3,4 etc. para-phenols.
In that case the terms of all the molecules have to be taken into account and average parameter value can be chosen to minimize the transformations.  
 
\subsection{Simulation Protocol}

We have used the Gromos53a6 force field parameters from ATB server
\cite{malde2011automated} for both p-Cresol and p-Chlorophenol (p-CH3
and p-Cl) along with spc water model during the simulations. The rhombic dodecahedron box,
with a minimum distance of 1nm between the solute (the p-CH3 or p-Cl molecule) and the
box edge to prevent interactions of the molecule with its periodic copy in the adjacent cell, was used. The rhombic dodecahedron was selected since it provides a
more effective packing of periodic images than rectangular boxes.
After minimization and equilibration (200 ps), 20 ns and 1 ns of MD simulations were performed in vacuum and water respectively under NVT and
NPT ensembles, at each of the 19 equi-spaced intermediate states
($\lambda$ states) including the initial ($\lambda=0$) and final states ($\lambda=1$). For each of the
$\lambda$ states, first the coulomb terms were relaxed followed by VDW terms.
We have computed the free energy difference between solvation of p-CH3 and p-Cl using the BAR
(Bennett's acceptance ratio) method \cite{bennett1976efficient}.

\subsection{Results and comparision}
We first calculated Helmholtz and Gibbs free energies corresponding to NVT and NPT ensembles respectively, according to the decoupling analysis in Sec. II. To this end we associated the bond angle of the C/Cl atom with the common subsystem and the the C-C/Cl-C atom distance with the different subsystem (the rest of the analysis follows from Sec. II). We then performed an additional transformation of p-CH3 and p-Cl molecules in which bond angle terms were removed according to the existing decoupling analysis. To perform the additional bonded transformation we defined for p-CH3 molecule the CH3 group atoms as the different subsystem and we removed the five bond angle terms and the dihedral angle term which involve atoms from the common and different subsystems. In the additional bonded transformation of p-Cl we defined Cl as the different subsystem and removed the two bond angle terms which involve atoms from the two subsystems. The simulations have been performed with bond length constraints that should have a minor effect on the dynamics. In the additional bonded transformation of p-Cl in vacuum in order to reduce the standard deviation we also performed in some of the intermediates a transformation without constraints.

The results obtained are the following (in kJ/mol):

\begin{widetext}

\begin{tabular}{|c|c|c|c|c|}
\hline 
 & \multicolumn{2}{c|}{NVT} & \multicolumn{2}{c|}{NPT}\tabularnewline
\hline 
 & vacuum & water & vacuum & water\tabularnewline
\hline 
\hline 
p-CH3 & 140.18\mbox{$\pm$}0.00 & 147.44\mbox{$\pm$}0.14 & 155.78\mbox{$\pm$}0.03 & 164.2\mbox{$\pm$}0.23\tabularnewline
\hline 
p-Cl & 19.74\mbox{$\pm$}0.01 &29.95\mbox{$\pm$}0.65 & 46.78\mbox{$\pm$}0.02 & 59.06\mbox{$\pm$}0.28\tabularnewline
\hline 
p-CH3 bonded terms & -25.51\mbox{$\pm$}0.10 &  -25.89\mbox{$\pm$}0.37 &  -25.72\mbox{$\pm$}0.25 & --26.87\mbox{$\pm$}0.22\tabularnewline
\hline 
p-Cl bonded terms & -7.68\mbox{$\pm$}0.78 & -7.51\mbox{$\pm$}0.24 &  -8.27\mbox{$\pm$}0.79 & -7.36\mbox{$\pm$}0.07\tabularnewline
\hline 
p-Cl bonded, no costraints & -7.42\mbox{$\pm$}0.25 &  &  -7.27\mbox{$\pm$}0.04 & \tabularnewline
\hline 
\end{tabular}
\end{widetext}

It can be seen that the free energy difference associated with the removal of the bonded terms is the same (within less than a standard deviation) in vacuum and water in NVT ensemble as explained in Section II. Gibbs free energy, which is usually similar to Helmholtz free energy \cite{abraham2015gromacs}, in vacuum and water was different by 1.15 kJ/mol for p-CH3 and the same for p-Cl.
In the following table we present Helmholtz and Gibbs relative free energies of solvation (in kJ/mol).

\medskip{}
\begin{tabular}{|c|c|c|c|}
\hline 
 & \begin{tabular}{@{}c@{}}decoupling \\ analysis\end{tabular} &\multicolumn{2}{c|}{inc. bonded} \tabularnewline
\hline 
& &constraints &no const.  \tabularnewline
\hline 
\hline 
\mbox{$\Delta F_{\mathrm{p-Cl\rightarrow p-CH_{3}}}$}  & -2.95\mbox{$\pm$}0.66  &  -3.5\mbox{$\pm$}1.12& -3.24\mbox{$\pm$}0.84 \tabularnewline
\hline 
\mbox{$\Delta G_{\mathrm{p-Cl\rightarrow p-CH_{3}}}$}  & -3.86\mbox{$\pm$}0.28  & -5.92\mbox{$\pm$}0.91 &-4.92\mbox{$\pm$}0.44 \tabularnewline
\hline 
\end{tabular}
\medskip{}

It can be seen that $\Delta F_{\mathrm{p-Cl\rightarrow p-CH_{3}}}$ is the same (within less than a standard deviation) when we removed additional bonded terms. Thus, a smaller transformation and a simpler implementation resulted in the same relative free energy with a shorter simulation and a smaller standard deviation. This implies that in order to obtain the same standard deviation using the existing decoupling analysis a longer simulation is required. 
 For similar transformations (in another context) see Ref. \cite{farhi2016calculation}.

%
%

The experimental free energy difference is \cite{rizzo2006estimation} :
$$\Delta G_{\mathrm{diff}}= -29.54 - (-25.75)
= -3.79\mathrm{ kJ/mol}.$$ 

It can be seen that there is good agreement between the calculation and the experimental free energies. The total simulation time using the decoupling analysis in Sec. II was $\sim42$ days (on a single core). Hence, improvement in the efficiency of the calculation is significant in terms of computation time.  
 
\section{Removing singularities at small $\lambda$s }
\label{section:soft_core}

The potential in the standard soft core technique is given by:
\begin{equation*}
H\left(\lambda,r\right)=4\epsilon\lambda^{n}\left[\left(\alpha\left(1-\lambda\right)^{m}+\left(\frac{r}{\sigma}\right)^{6}\right)^{-2}\right.+ 
\end{equation*}
$$\left.\left(\alpha\left(1-\lambda\right)^{m}+\left(\frac{r}{\sigma}\right)^{6}\right)^{-1}\right].$$

Here we present a soft core technique in which we cap the diverging energy terms at high energetic values in order to remove the singularities at small $\lambda$s. This is a unification of the approach in Ref. \cite{FarhiThesis} in which  accessible capping energies with a negligible effect on the free energy are suggested, resulting in integrated functions that are less steep, and Ref. \cite{buelens2012linear} in which the derivative of the potential is continuous, enabling use in MD simulations. We also show mathematically that that the monotonicity of the integrated function is ensured.

This soft core technique does not introduce dependency on $\lambda$ and hence the potential and its derivative are relatively simple to implement. In addition, the original shape of the potential constant, which is good in terms of phase space overlap. Moreover, the need to remove first the electrostatic terms and then the VDW terms to avoid singularities is eliminated.


\subsection {The value of the capping energy}
\label{subsec:capping_energy_value}
Since at $\lambda=0$ the energy terms
 that diverge at $r=0$ cause the average energy to diverge,
capping is used in the long range energy terms (if $E>E_{\mathrm{cap}}$, $E=E_{\mathrm{cap}}$).
Thus, the terms at $\lambda\rightarrow0$ are no longer dominant and decoupling is achieved.
The proposed calculation of the free energy difference between the
two systems is legitimate only if the choice of the capping energy
has a negligible effect on the free energy value of each of
the two systems at $\lambda=1$. The Hamiltonian with the capped
 long range energy terms 
$H'$ is written as follows:
\begin{equation} \label{eq:separate}
H'_{A/B}\left(\beta,\lambda\right)=\lambda
H'_{_{{A_{\mathrm{lr}}/B_{\mathrm{lr}}}}}+H_{{}{A_{\mathrm{sr}}/B_{\mathrm{sr}}}},
\end{equation}
where $H'_{_{{A_{\mathrm{lr}}/B_{\mathrm{lr}}}}}$ and $H_{{}{A_{\mathrm{sr}}/B_{\mathrm{sr}}}}$ denote the capped long range and the short range terms respectively.
The requirement stated above can be written explicitly as follows:
\begin{equation}\label{cutoffA}
\ln {Z}_{A/B}\left(\beta,\lambda=1,H'\right)\backsimeq
\ln {Z}_{A/B}\left(\beta,\lambda=1,H\right).
\end{equation}


In order for the capping to have a negligible effect on the partition
functions at $\lambda=1$ it has to be set to a value that satisfies:
\begin{equation}
e^{-\frac{E_{\mathrm{cap}}}{kT}}\ll 1 .
\end{equation}
Thus at $\lambda$ values satisfying $e^{-\frac{\lambda E_{\mathrm{cap}}}{kT}}\approx1$ the diverging interactions, including the steric,
become transparent.
\begin{figure}[h]
 \centering
\includegraphics[width=8cm]{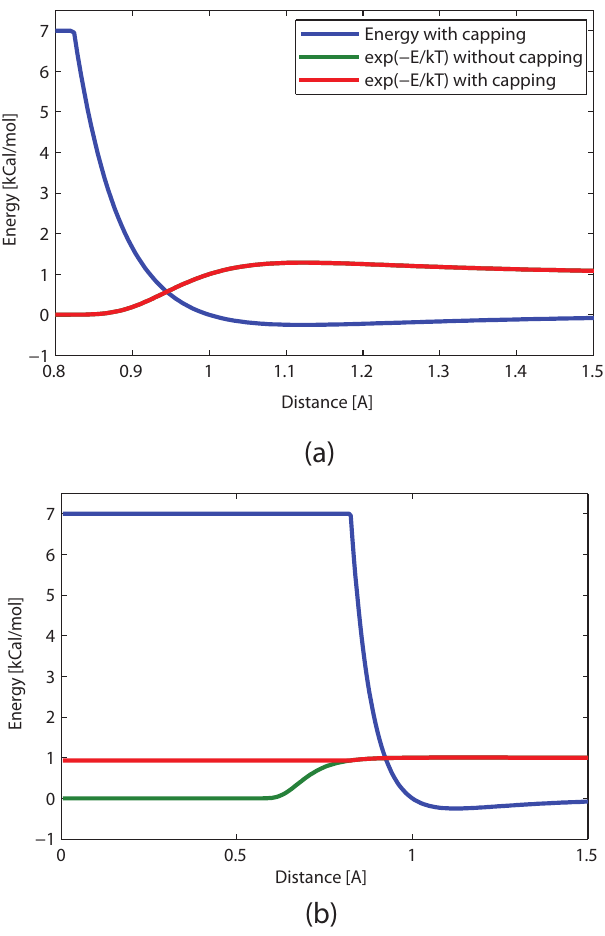}
 \caption{Energy and $\exp(-E/kT)$ as a function of distance for the potential $r^{-12}-r^{-6}$ with $E_{\mathrm{cap}}$=7kcal/mol (a) at $\lambda=1$ (b)  at $\lambda=0.01$}
 \label{fig:cutoff}
\end{figure}
In Fig. \ref{fig:cutoff} energy and $\exp(-E/kT)$ as a function of distance for the potential $r^{-12}-r^{-6}$ are plotted at $\lambda=1$ and at $\lambda=0.01$. It can be seen that the capping of the energy has a negligible effect on the probability distribution at $\lambda=1$ and that at small $\lambda$s the interactions are transparent.

It is suggested that since the probability to be in a microstate decays exponentially with the energy with typical decay scale of $k_{B}T$ and since the density of states for $E>E_{\mathrm{cap}}$ is very low, capping the energy at a value several times $k_{B}T$ higher than the equilibrium total atom-atom long range energy will have a negligible effect on the free energy value at $\lambda=1$.

\subsubsection*{Demonstrations}

It has been demonstrated that $E_{\mathrm{cap}}$ values of $\thicksim5$kcal/mol enable accurate free energy calculations \cite{FarhiThesis}. In this reference the free energy calculation results  for $E_{\mathrm{cap}}=5$kcal/mol and $E_{\mathrm{cap}}=4.4$kcal/mol were similar (see Fig. 29 there). In addition in Ref. \cite{farhi2013general} the free energy difference calculation between two systems composed of two atoms with $E_{\mathrm{cap}}=7$kcal/mol was very accurate (see Supplementary Material which appears in the same document). See also Fig. \ref{fig:cutoff}. Note that in Ref. \cite{FarhiThesis} $\beta$ is varied rather than $\lambda$. However, the soft core technique presented there is mathematically equivalent to the one presented here. The conditions for the capping energy here involve $\lambda$ rather than $\beta$.

\subsection{The monotonicity of the integrated function}
In this subsection we explain why the suggested soft core technique ensures the monotonicity of the integrated function.
 This monotonicity will enable a simple selection of intermediates for the calculation of the free
energy difference. For example of integrated functions that are not monotonic see Ref. \cite{khavrutskii2010computing}, Fig. 3. 

This monotonicity for the soft core technique when used with the transformation of Eq. (\ref{eq:second_transformation}) can be understood by recalling that the integrated function $\left\langle H_{A_r}\right\rangle$ in Eq. (\ref{eq:integration})  is calculated with the governing Hamiltonian $ H_{A_c}+\lambda H_{A_r}$.
 A system will spend time in each configuration proportional to the Boltzmann weight of that configuration which is determined by the governing Hamiltonian. When energy terms are multiplied by a $\lambda$ value smaller than 1, the energy heights and valleys of that terms will appear smaller. Thus, the system will spend more time in these heights and less time in these valleys. In the calculation of the average energy, the energy landscape did not change as the Hamiltonian remained the same. Thus, since the system now spends more time in less favorable states, the calculated average energy will be higher. This is similar to heating the system as $\lambda$ plays the role of $\beta$. Thus low values of $\lambda$ correspond to high temperatures, leading to higher average energies. This is expressed in the known result  $-\frac{\partial U}{\partial\beta}=\left\langle \left(\Delta E\right)^{2}\right\rangle 
 $  \cite{pathriastatistical}.   However, here $\lambda$ multiplies \emph{some} of the terms so it can be regarded as ``partial heating'' (see Eq. (\ref{eq:second_transformation})). Also, in this case the average energy $\langle H_r \rangle$  is only of the removed terms.

We now prove that the integrated is monotonic for transformations which depend linearly on $\lambda$.
It is easy to show that for $\frac{d^{2}H\left(\lambda\right)}{d\lambda^{2}}=0$ (see detailed proof in Appendix \ref{Appendix:monotonicity_proof})
\begin{equation}
-\frac{d}{d\lambda}\left(\frac{dF}{d\lambda}\right)=\beta\left\langle \left[\frac{dH\left(\lambda\right)}{d\lambda}-\left\langle \frac{dH\left(\lambda\right)}{d\lambda}\right\rangle \right]^{2}\right\rangle =\beta\sigma^{2}\geq 0, 
\end{equation}
where $\sigma^{2}=\left\langle \left[\frac{dH\left(\lambda\right)}{d\lambda}-\left\langle \frac{dH\left(\lambda\right)}{d\lambda}\right\rangle \right]^{2}\right\rangle$ . 

Hence, the suggested soft core technique, which does not introduce dependency on $\lambda$ combined with linear transformations in $\lambda$ such as the transformations of Eq. (\ref{eq:first_transformation}) or Eq. (\ref{eq:second_transformation})
 results in a monotonic change in the integrated function. 

Interestingly, this also shows that the variance of $\frac{dH\left(\lambda\right)}{d\lambda}$  is proportional to the slope of the integrated function. Thus, steep slopes of the integrated function are challenging both to sample and to numerically integrate. 

The monotonicity is important for the integration since at two adjacent intermediates there can be no extremum point, which can cause a free energy difference that is not taken into account in the numerical integration.  The monotonicity of the function enables accurate numerical integration and facilitates the selection of intermediates.

 Setting $\lambda\rightarrow\lambda^{n}$  preserves monotonicity in both hybrid and separate transformations. When performing separate transformations there is a ``soft core transition'' (steep integrated function) near $\lambda=0$. When performing hybrid transformations there are two ``soft core transitions'' - near $\lambda=0$  and near $\lambda=1.$  Setting $\lambda \rightarrow\lambda^n$ when $n>1$
  results in a less steep integrated function near $\lambda=0$  and more steep integrated function near $\lambda=1$. This is useful in separate transformations
  but not in hybrid transformation in which it will improve the sampling for one transition but worsen for the other transition. (see Appendix \ref{Appendix:improved_behavior} for more details).

This statement regarding the monotonicity of the integrated function is in agreement with Ref. \cite{buelens2012linear} (Fig. 5), in which the soft core technique (for inaccessible capping energy) with a linear transformation of the type of Eq. (\ref{eq:second_transformation}) results in a monotonic change in the integrated function (has not been pointed out).   

\subsection{The effect of the capping energy on the integrated function}

We now analyze the effect of the capping energy value on the integrated function. We choose two legitimate capping energies  $E_{\mathrm{cap}}$=7kcal/mol and $E_{\mathrm{cap}}$=15kcal/mol. For simplicity in the following paragraph we omit the units. 
The free energy change associated with the transition to "transparent" VDW and electrostatic interactions is concentrated in the range $0.05\lesssim\exp\left(-\beta\lambda E_{\mathrm{cap}}\right)\lesssim0.95.$ 
Taking $\beta\rightarrow1$  for simplicity, we get that the free energy change is concentrated in $\frac{0.05}{E_{\mathrm{cap}}}\lesssim\lambda\lesssim\frac{3}{E_{\mathrm{cap}}}.$ This translates into $0.007\lesssim\lambda\lesssim0.42$  and $0.003\lesssim\lambda\lesssim0.2$  for $E_{\mathrm{cap}}=7$ and $E_{\mathrm{cap}}=15$  respectively. It can thus be seen that the range in which the free energy changes is smaller and the change occurs at smaller $\lambda$s as $E_{\mathrm{cap}}$  is higher. Since $\intop_{0}^{1}\left\langle \frac{dH_{A}}{d\lambda}\right\rangle d\lambda$ is the same to a good accuracy for both capping energies (equal area), the integrated function reaches a higher value at $\lambda=0$ for high $E_{\mathrm{cap}}$  values.

\subsubsection*{Demonstrations}
It has been demonstrated in molecular MC simulations that there is a trade-off when choosing $E_{\mathrm{cap}}$. That is, high $E_{\mathrm{cap}}$ value results in a more accurate calculation but an integrated function that is steeper and reaches a higher value. See Ref. \cite{FarhiThesis} (Fig. 28) in which the integrated functions for $E_{\mathrm{cap}}=5$kcal/mol and $E_{\mathrm{cap}}=4.4$kcal/mol are compared. See also the inset in Fig. 5 in Ref. \cite{buelens2012linear} in which the integrated function with $E_{\mathrm{cap}}=40$kcal/mol is plotted (MD simulations). 

\subsection{Continuity of the potential and its derivative}
In order to have continuity in the derivative that will enable the integration over the equations of motion in MD to be valid, a switching function between the standard long range potential and the flat potential is needed. This has been developed independently and implemented in MD with a cubic 
switching function and an energetically inaccessible capping energy (reaches a state in which $E>E_{\mathrm{cap}}$ once every $10^{14}$ moves), which validates the use of the capping in the context of MD for high energetic values (40kcal/mol) \cite{buelens2012linear}. This use of a switching function with $E_{\mathrm{cap}}=40$kcal/mol as a soft core techniques appeared to perform marginally better compared to the standard soft core technique \cite{buelens2012linear}.

\subsection{Unifying the approaches}
 Thus a unified approach that uses a switching function and a capping energy that is accessible and has a negligible effect on the free energy (e.g $E_{\mathrm{cap}}=7$kcal/mol) is suggested as a soft core technique. This condition results in significantly lower value of $E_{\mathrm{cap}}$ as compared with the one needed in order to ensure inaccessibility and thus sampling is much easier (see the comparison of the behavior of the integrated function for two different $E_{\mathrm{cap}}$ values in Ref. \cite{FarhiThesis} Fig. 28).
Now Eq. (\ref{eq:relative_free_energies}) can be written as follows:
\begin{equation*}
\Delta F_{A_{\mathrm{solvation/binding}}\rightarrow B_{\mathrm{solvation/binding}}}=
\end{equation*}
\begin{equation*}
\int_0^1\left\langle H'_{B_{r}}\right\rangle -\int_0^1\left\langle H'_{A_{r}}\right\rangle 
\end{equation*}
\begin{equation}
+\int_0^1\left\langle H'_{A_{\mathrm{solvated/bounded}_{r}}}\right\rangle -\int_0^1\left\langle H'_{B_{\mathrm{solvated/bounded}_{r}}}\right\rangle.
\end{equation}
%

\section{Sampling rugged energy landscape in one $\lambda$ dimension}
\label{section:one_dimension}
When the systems, between which the free energy difference is
calculated, have rugged energy landscape in conformational space (as a function of the coordinates of the atoms), one can use techniques
such as H-REMD/H-PT (Hamiltonian Replica Exchange MD/ Hamiltonian Parallel Tempering, variant of Parallel Tempering/Replica Exchange \cite{ferrenberg1988new,hansmann1997parallel,earl2005parallel}) to alleviate sampling problems \cite{fukunishi2002hamiltonian}. In this technique the system is simulated at a set of $\lambda$s and
exchanges of configurations between them are performed every
certain number of steps according to the Metropolis criterion. 
Thus, the systems at the low $\lambda$s,
that can cross energetic barriers, help the system of interest to
be sampled well .This technique, even though is highly efficient,
introduces another sampling dimension since the simulations of the
replicas at a set of $\lambda$s are performed at
each intermediate of the hybrid system (sampling the dimensions of
$\lambda$ that interpolates between the systems and of $\lambda$
of the replicas that are used for the equilibration). Here, the
simulations at the different $\lambda$s will be used \emph{also}
to calculate the free energy difference by integration and
the need for another sampling dimension is eliminated.
 
\begin{figure}[h]
 \centering
\includegraphics[width=8cm]{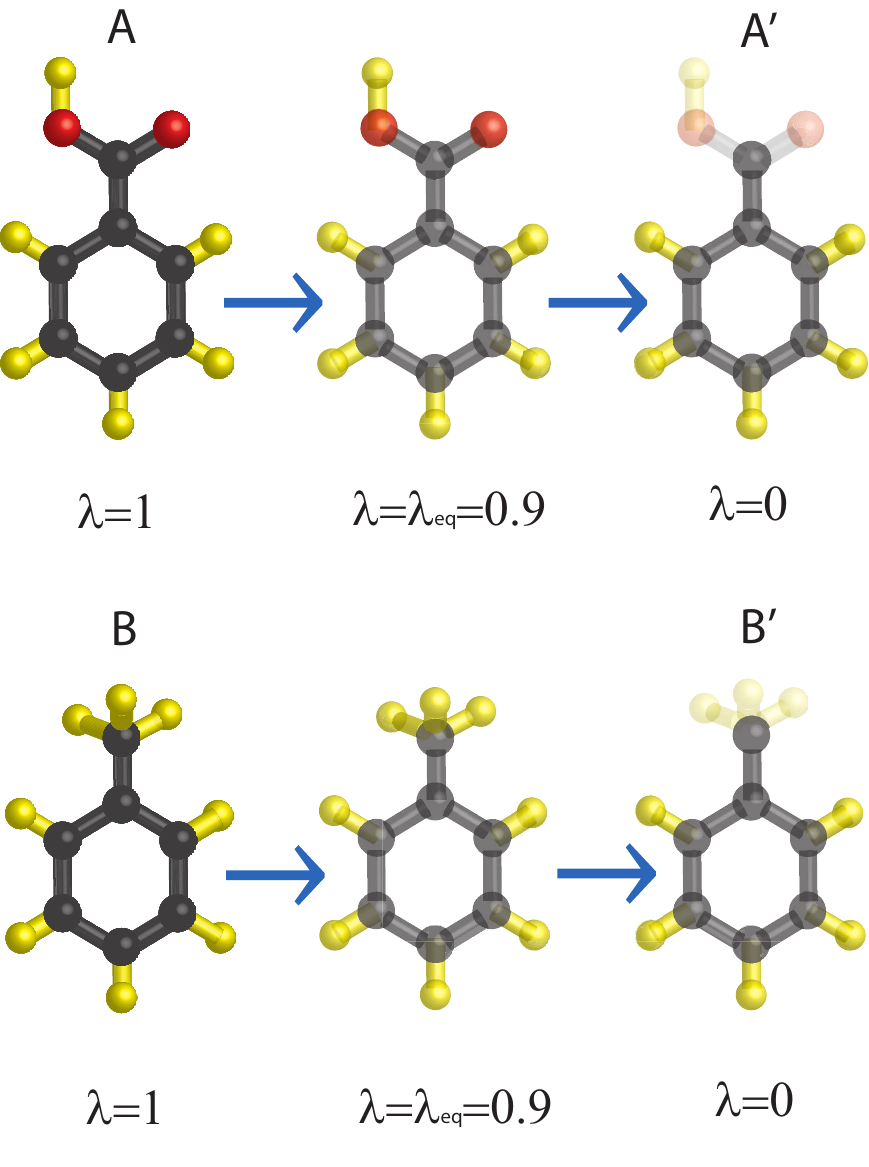}
 \caption{A scheme of the two transformations suggested. At $\lambda=\lambda_{\mathrm{eq}}$ all the energy terms that need equilibration are multiplied by $\lambda_{\mathrm{eq}}$. At $\lambda=0$ $H_r$ is removed and the other terms remain multiplied by the same $\lambda$ as when $\lambda=\lambda_{\mathrm{eq}}$ .}
 \label{fig:one_sampling_dimension}
\end{figure}

In order to equilibrate the entire system, the energy terms that are not multiplied by $\lambda$ can be written as follows:
\begin{equation}
H_{c}\rightarrow f\left(\lambda \right)H_{c},\,\,\,\,\, f\left(\lambda\right)=\left\{ \begin{array}{cc}
\lambda & \lambda\geq\lambda_{\mathrm{eq}},\\
\lambda_{\mathrm{eq}} & \lambda<\lambda_{\mathrm{eq}},
\end{array}\right.
 \end{equation}
where $\lambda_{\mathrm{eq}}$ denotes the minimal $\lambda$ for equilibration in the H-REMD procedure.
Here we transform all the terms only up to $\lambda_{\mathrm{eq}}$ in order to have a minimal transformation.
Thus the H-REMD procedure is in its original form in the range $\lambda=[1,\lambda_{\mathrm{eq}}]$ and the systems at $\lambda=[\lambda_{\mathrm{eq}},0]$, in which only $H_r$ is lowered, can be simulated separately since the energy barriers are accessible for these $\lambda$ values.
 See Fig. \ref{fig:one_sampling_dimension}.

H-REMD, which is in its standard use here, is explained and demonstrated in Ref. \cite{fukunishi2002hamiltonian}. A general example which demonstrates free energy calculation of systems with rugged energy landscape in one $\lambda$ dimension with further methodological advances can be seen in Ref. \cite{farhi2013general}, Supplementary Material. For methodologies and guiding principles in choosing the intermediates in the H-REMD/PT procedure see Refs. \cite{trebst2006optimized,nadler2008optimized,FarhiTemperature}.

\section{Discussion}
\label{section:discussion}
A novel method for calculating relative free energies is presented. The method can be used to calculate the free energy difference between solvation/binding free energy of two molecules with any number of atoms and is applicable to MD and MC simulations and  to all types of molecular modelings. The article is composed of several independent ingredients. First, we showed that the partition functions of the common and different subsystems decouple exactly and hence there is no error involved. This analysis is applicable to all potential functions and to submolecules with coupled degrees of freedom. Then we suggested to use the two separate systems instead of one system that includes ingredients of the two systems in order to calculate the relative free energy. This has the advantage of large phase space overlap since the systems are inherently correlated and \emph{simplicity} since the simulations are performed only on the two (almost) original systems in two separate simulations and the need for extensive design is eliminated. The third ingredient is a unified approach to soft core potentials.
 This technique is simple to implement and results in monotonic change of the integrated function. The monotonicity enables \emph{simple} selection of intermediates and ensures accuracy in the numerical integration and hence a \emph{robust} result.

We also show how if the systems have rugged energy landscape, instead of using the sampling techniques in another $\lambda$ or $T$ dimension, we can use only one sampling dimension.  since the $\lambda$s used in the H-REMD/H-PT procedure are also used as intermediates in the calculation of free energy difference, a
convergence for systems with rugged energy landscape is achieved
without introducing another sampling dimension. Both in the
calculation of the integral for the free energy difference and in
the H-REMD/H-PT procedures, the chosen intervals between the
$\lambda$s have to be smaller where the internal energy varies
significantly, in the free energy difference calculation in order
to have good sampling of the function and in H-REMD in order to
maintain optimal acceptance rates. Thus, no additional unnecessary
$\lambda$s have to be sampled.

It has been shown analytically and in MD simulations that less terms need to be removed in the transformation as compared to the existing methodology such as the dihedral terms. Since each removal of a term in the transformation has a free energy energy value associated with it, removing less terms necessarily means that the free energy difference in the transformation is smaller and less intermediate systems are required. In addition the monotonicity of the integrated function in the soft core technique has both been proved analytically and backed up in existing MD simulations. The monotonicity of the integrated function ensures that the function will not have extrema and enables to know the upper limit of the numerical integration error. Thus it is expected that less intermediates will be needed to sample the function. The relation between the free energy difference and the number of intermediates required can be understood for example from the upper limit of the numerical integration error of a monotonic function $\Delta x\Delta y/2$ which depends linearly on this difference. Thus, for integration result within a given numerical error, less intermediates are expected to be required.  Finally, capping the potential at an energy value that is energetically accessible in the soft core technique, which has been demonstrated in MC simulations, results in a function that is less steep and reaches a lower value as compared to the case of an energetically inaccessible capping energy. Thus, again, less intermediates are required. Since the number of intermediates is directly related to the computational power needed, the principles presented here are expected to increase the \emph{efficiency} of the calculations.   

These advantages are of high importance for automating free energy calculations and computational drug discovery. Thus, using this method, preceded by virtual
screening filtering, an automated free energy calculation
that will result in the best candidates may be
 performed. It is noted that the method may have other applications in physics, where the environment for example can be external electric or magnetic field.

\appendix

\section{Molecular potentials}
\label{Appendix:potentials}

Here we  briefly describe the molecular potentials.

The covalent bond between two atoms is modeled with the following potential:
\begin{equation}
V_{c}=k_{c}\left(r-d\right)^{2},
\end{equation}
where $r$ is the distance between the atoms.
The bond angle term between three atoms is usually modeled with the following potential:
\begin{equation}
V_{b}\left(\theta\right)=\frac{1}{2}k_{\theta}\left(\theta-\theta_{0}\right)^{2}.\end{equation}
The commonly used dihedral angles potential is of the following type:
\begin{equation}
V_{d}\left(\phi_{ijkl}\right)=k_{\phi}\left(1+\cos\left(n\phi_{ijkl}-\phi_{s}\right)\right),\end{equation}
where $\phi_{ijkl}$ is defined by the angle between the plane formed by the $ijk$ atoms and the plane formed by the $jkl$ atoms.
The improper dihedral term that is used to enforce planarity is defined as follows:
\begin{equation}
V\left(\phi_{ijkm}\right)=\frac{1}{2}k\left(\phi_{ijkm}-\phi_{s}\right)^{2},\end{equation}
where $\phi_{ijkm}$ is defined by the angle between the plane formed by the $ijm$ atoms and the plane formed by the $imk$ atoms. The atoms in this case have different covalent connectivity.
\begin{figure}[htp]
 \centering
\includegraphics[width=4cm]{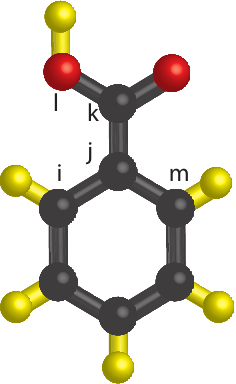}
 \caption{Benzoic Acid molecule with atom indices that suit the defined dihedral terms}
 \label{fig:potentials}
\end{figure}
The Coulomb and VDW terms are defined by the $1/r$ and $r^{-12}-r^{-6}$ potentials respectively. The coefficients for these interactions are determined by the VDW coefficients/charges of both atoms. 

\section{Decoupling the partition function - detailed proof}
\label{appendix:detailed_proof}
Since we vary over all possible values of $\mathbf{\Omega}_{\mathrm{dif}}$
  the integration result does not depend on $\hat{x}$
  and $\hat{z}$.

We can change the coordinate system of the different subsystem so that $\hat{x}$
  and $\hat{z}$
  and accordingly $\hat{y}$
  will be its new axes $\hat{x}_{\mathrm{new}}=O^{-1}x_{\left(0,0,1\right)}=\hat{x}$
  and similarly for $y$
  and $z$
  axes, where $O^{-1}$
  is the rotation matrix that rotates the previously defined axes $\left[\left(0,0,1\right),\left(0,1,0\right),\left(1,0,0\right)\right]$
  to $\hat{x},\hat{y},\hat{z}
 .$ $\mathbf{\Omega}_{\mathrm{dif}}$
  is rotated with $O$
  in order to be represented in the new system of coordinates $\mathbf{\tilde{\Omega}}_{\mathrm{dif}}=O\mathbf{\Omega}_{\mathrm{dif}}.$
  The Jacobian of rotation is 1 $d\mathbf{\tilde{\Omega}}_{\mathrm{dif}}=d\mathbf{\Omega}_{\mathrm{dif}}.$
  The integration limits of $\tilde{x}_{j},\tilde{y}_{j}$
  and $\tilde{z}_{j}$
  are $-\infty$
  to $\infty.$ $\hat{x}_{\mathrm{new}}$
  and $\hat{z}_{\mathrm{new}}$
  in the new system of coordinates are $\left(0,0,1\right)$
  and $\left(1,0,0\right).$ We can thus write

$$\int e^{-\beta H_{\mathrm{com}}\left(\mathbf{\Omega}_{\mathrm{com\,}}\right)}d\mathbf{\Omega}_{\mathrm{com}}\int e^{-\beta H_{\mathrm{dif}}\left(\tilde{\mathbf{\Omega}}_{\mathrm{dif}},\left(0,0,1\right),\left(1,0,0\right)\right)}d\mathbf{\tilde{\Omega}}_{\mathrm{dif}}=$$
 
$$
\int e^{-\beta H_{\mathrm{com}}\left(\mathbf{\Omega}_{\mathrm{com\,}}\right)}d\mathbf{\Omega}_{\mathrm{com}}Z_{\mathrm{dif}}=Z_{\mathrm{com}}Z_{\mathrm{dif}}.$$
 
Alternatively, we can switch $\mathbf{\Omega}_{\mathrm{dif}}$
  to relative spherical variables and this integration will not depend on $\mathbf{\Omega}_{\mathrm{com\,}}$.
  Here we define $\hat{x}$,$\hat{z}$
  for every atom in the different submolecule similarly to the definitions above. We can thus write 

$$Z=\int e^{-\beta H_{\mathrm{com}}\left(\mathbf{\Omega}_{\mathrm{com\,}}\right)}d\mathbf{\Omega}_{\mathrm{com}}\times $$ $$ \int e^{-\beta H_{\mathrm{dif}}\left(r_{k+1},r_{k+2},r_{k+3},\theta_{k+1},\theta_{k+2},\theta_{k+3},\phi_{k+1},\phi_{k+2},\phi_{k+3}\right)}\times$$ $$\prod_{j=k+1}^{k+3}r_{j}^{2}\sin\theta_{j}dr_{j}d\theta_{j}d\phi_{j}=$$

$$\int e^{-\beta H_{\mathrm{com}}\left(\mathbf{\Omega}_{\mathrm{com\,}}\right)}d\mathbf{\Omega}_{\mathrm{com}}Z_{\mathrm{dif}}=Z_{\mathrm{com}}Z_{\mathrm{dif}}$$
 
\section{Detailed proof of the monotonicity of the integrated function}
\label{Appendix:monotonicity_proof}

\begin{equation}
-\frac{d}{d\lambda}\left(\frac{dF}{d\lambda}\right)=\frac{\int e^{-\beta H\left(\lambda\right)}\left[\beta\left(\frac{dH\left(\lambda\right)}{d\lambda}\right)^{2}-\frac{d^{2}H\left(\lambda\right)}{d\lambda^{2}}\right]d\Omega}{Z\left(\lambda\right)}\nonumber
\end{equation}
\begin{equation}
-\int e^{-\beta H\left(\lambda\right)}\frac{dH\left(\lambda\right)}{d\lambda}d\Omega\frac{d\left[Z\left(\lambda\right)^{-1}\right]}{d\lambda}. 
\end{equation}
Differentiating $Z\left(\lambda\right)^{-1}$  with respect to $\lambda$ we get:
\begin{equation}
\frac{d\left[Z\left(\lambda\right)^{-1}\right]}{d\lambda}=-Z\left(\lambda\right)^{-2}\frac{dZ}{d\lambda}=-Z\left(\lambda\right)^{-2}\frac{d\int e^{-\beta H\left(\lambda\right)}d\Omega}{d\lambda}=\nonumber
\end{equation}
\begin{equation}
Z\left(\lambda\right)^{-2}\int e^{-\beta H\left(\lambda\right)}\beta\frac{dH\left(\lambda\right)}{d\lambda}d\Omega.\nonumber 
\end{equation}
And finally:
\begin{equation}
-\frac{d}{d\lambda}\left(\frac{dF}{d\lambda}\right)=\frac{\int e^{-\beta H\left(\lambda\right)}\left[\beta\left(\frac{dH\left(\lambda\right)}{d\lambda}\right)^{2}-\frac{d^{2}H\left(\lambda\right)}{d\lambda^{2}}\right]d\Omega}{Z}\nonumber
\end{equation}
\begin{equation}
-\beta\left\langle \frac{dH\left(\lambda\right)}{d\lambda}\right\rangle \left\langle \frac{dH\left(\lambda\right)}{d\lambda}\right\rangle . 
\end{equation}
For $H$  that depends linearly on $\lambda$ the second derivative vanishes and we can write:
\begin{equation}
-\frac{d}{d\lambda}\left(\frac{dF}{d\lambda}\right)=\beta\left[\left\langle \left(\frac{dH\left(\lambda\right)}{d\lambda}\right)^{2}\right\rangle -\left\langle \frac{dH\left(\lambda\right)}{d\lambda}\right\rangle \left\langle \frac{dH\left(\lambda\right)}{d\lambda}\right\rangle \right]=\nonumber
\end{equation}
\begin{equation}
\beta\left\langle \left[\frac{dH\left(\lambda\right)}{d\lambda}-\left\langle \frac{dH\left(\lambda\right)}{d\lambda}\right\rangle \right]^{2}\right\rangle =\beta\sigma^{2}\geq 0, 
\end{equation}
where $\sigma^{2}=\left\langle \left[\frac{dH\left(\lambda\right)}{d\lambda}-\left\langle \frac{dH\left(\lambda\right)}{d\lambda}\right\rangle \right]^{2}\right\rangle$ .

\section{Setting $\lambda\rightarrow\lambda^{n}$ preserves monotonicity and results in improved behavior of the integrated function when used in separate transformations}
\label{Appendix:improved_behavior}
We now explain that when performing separate simulations we can choose a transformation that will preserve the monotonicity and result in improved behavior of the integrated function. We will investigate free energy functions which can be mapped from the free energy functions with linear transformations. For example the free energy for the separate and linear transformation $H_{ls}\left(\lambda\right)=\lambda H_{\mathrm{r}}+H_{\mathrm{c}}$
  can be mapped into this of the transformation $H_{ms}\left(\lambda\right)=\lambda^{n}H_{\mathrm{r}}+H_{\mathrm{c}}$
  ($F_{ls}\left(\lambda^{n}\right)=F_{ms}\left(\lambda\right)$).
  The integrated function of a mapped function $F_{m}\left(\lambda\right)$
  from a linear function $F_{l}\left(\lambda\right)$
  for any $\lambda\rightarrow\lambda^{n},\lambda\in\left[0,1\right]$
  mapping is monotonic

$$-\frac{d^{2}F_{m}\left(\lambda\right)}{d\lambda^{2}}=-\frac{d^{2}F_{l}\left(\lambda^{n}\right)}{d\lambda^{2}}=$$ $$-\frac{d^{2}F_{l}\left(\lambda^{n}\right)}{\left(d\lambda^{n}\right)^{2}}\left(\frac{d\lambda^{n}}{d\lambda}\right)^{2}=-\frac{d^{2}F_{l}\left(\lambda\right)}{d\lambda^{2}}\left(n\lambda^{n-1}\right)^{2}>0.$$

It can be readily seen that the hybrid and linear transformation $H_{lh}\left(\lambda\right)=\lambda H_{\mathrm{r}}+\left(1-\lambda\right)H_{\mathrm{c}}$
  cannot be mapped into the transformation $H_{mh1}\left(\lambda\right)=\lambda^{n}H_{\mathrm{r}}+\left(1-\lambda\right)^{2}H_{\mathrm{c}}$.
  For example for $n=2,\,\lambda=0.1$
  we have $H_{mh1}\left(\lambda=0.1\right)=0.01\cdot H_{\mathrm{r}}+0.81\cdot H_{\mathrm{c}}$
  which has no corresponding value in $F_{lh}$. $F_{lh}$
  can be mapped into $H_{mh2}=\lambda^{n}H_{\mathrm{r}}+\left(1-\lambda^{n}\right)H_{\mathrm{c}}$.

Mapping of $\lambda\rightarrow\lambda^{n}$
  results in $\left(n\lambda^{n-1}\right)^{2}$
  factor which equals 1 at $\lambda=\sqrt[n-1]{\frac{1}{n}}$
  and for $n>1$  increases with $\lambda$. For example for $n=2$
  this factor equals 1 at $\lambda=\frac{1}{\sqrt{2}}$ . This means that the slope of the integrated function will be smaller for $\lambda<\frac{1}{\sqrt{2}}$
  and larger for $\lambda>\frac{1}{\sqrt{2}}$
  compared to the original one. Such a mapping will result in less steep integrated functions for transitions at small $\lambda$
 s but steeper integrated functions for transitions at high $\lambda$ s. Thus $\frac{dF_{ls}}{d\lambda}$
  in which the transition occurs at small $\lambda$
s will benefit from such a mapping but $\frac{dF_{lh}}{d\lambda}$
  which includes a transition near $\lambda\backsimeq1 $
  will experience steeper changes in the region of this transition (see for example \cite{farhi2013general} Fig. 3 (a)). In conclusion, $\lambda\rightarrow\lambda^{n},n>1$
  mapping in separate simulations will result in both monotonic integrated function and more equally spaced intermediates. 

\section{Sampling rugged energy landscape in one $\lambda$ dimension - additional explanations}

The free energy difference calculated in the H-REMD procedure, which is in the range $\lambda=[1,\lambda_{\mathrm{eq}}]$ can also be used for comparisons to other molecules that have a subsystem in common. The simulations in the range $\lambda=[\lambda_{\mathrm{eq}},0]$ can be used for comparison to other molecules that have the same common subsystem as the one in the transformation that was performed.
Covalent bond and bond angle energy terms may not need equilibration (multiplication by $\lambda$) as they are not expected to be associated with rugged energy landscape. It is emphasized that any transformation can be combined with H-REMD. However, it usually does not ensure convergence of the simulations and another dimension is needed. Since here we use only the original molecule in the transformation it can be performed in such a way that will ensure convergence. This originates from the fact that in order to achieve convergence the system needs to exchange configurations with its replica in a Hamiltonian with lowered energy barriers. Thus only when we relax terms up to a value in which the states can be sampled well will the simulations converge. This option is thus possible only when using the topology and soft core ingredients presented before. 

\subsection*{Acknowledgments}

D. J. Bergman, A. Nitzan, G. Falkovich, D. Harries, and R. Peled are acknowledged for the useful comments. I. Khavrutskii, M. Shirts, K. Vanommeslaeghe, F. Buelens and C. Junghans are acknowledged for the fruitful discussions.

\bibliographystyle{apsrev4-1}
%


\end{document}